\documentclass{aa}
\usepackage{natbib,psfig,graphicx,ulem}
\usepackage{txfonts}

\def\mathnew{\mathsurround=0pt}
\def\simov#1#2{\lower .5pt\vbox{\baselineskip0pt \lineskip-.5pt
\ialign{$\mathnew#1\hfil##\hfil$\crcr#2\crcr\sim\crcr}}}

\def\MeV{Me\kern-0.11em V}
\def\keV{ke\kern-0.11em V}

\begin{document}

\title{Spatial variations of the optical galaxy luminosity functions and 
red sequences in the Coma cluster: clues to its assembly history
\thanks{Based on observations obtained at the Canada-France-Hawaii Telescope 
(CFHT) which is operated by the National Research Council of Canada, the 
Institut National des Sciences de l'Univers of the Centre National de la 
Recherche Scientifique of France, and the University of Hawaii.}}

\author{C. Adami\inst{1} \and
F. Durret\inst{2,3} \and
A. Mazure\inst{1} \and
R. Pell\'o\inst{4} \and
J.P. Picat\inst{4} \and
M. West\inst{5} \and
B. Meneux\inst{1,6,7}
 }

\offprints{C. Adami \email{christophe.adami@oamp.fr}}

\institute{
LAM, Traverse du Siphon, 13012 Marseille, France
\and
Institut d'Astrophysique de Paris, CNRS, Universit\'e Pierre et Marie Curie,
98bis Bd Arago, 75014 Paris, France
\and
Observatoire de Paris, LERMA, 61 Av. de l'Observatoire, 75014 Paris, France
\and
Observatoire Midi-Pyr\'en\'ees, 14 Av. Edouard Belin, 31400 Toulouse, France
\and
Department of Physics and Astronomy,  University of Hawaii,
200 West Kawili Street, LS2, Hilo HI 96720-4091, USA
\and
INAF - IASF, Via Bassini 15, I-20133 Milano, Italy 
\and
INAF, Osservatorio Astronomico di Brera, Via Bianchi 46,
I-23807 Merate (LC), Italy 
}

\date{Accepted . Received ; Draft printed: \today}

\authorrunning{Adami et al.}

\titlerunning{Spatial variations of the Coma Luminosity Functions and 
Colour-Magnitude Relations}

\abstract
{Clusters of galaxies are believed to be at the intersections of
cosmological filaments and to grow by accreting matter from these
filaments.  Such continuous infall has major consequences not only on
clusters but also on the physics of cluster galaxies. Faint galaxies
are particularly interesting as they are very sensitive to
environmental effects, and may have a different behaviour from that of
bright galaxies.}
{The aim of this paper is to sample the Coma cluster building
history, based on the analysis of galaxy luminosity functions (LFs)
and red sequences (RSs) in the Color Magnitude Relation (CMR) down to
faint magnitudes, which are privileged tools for this purpose.}
{The present analysis is based on deep (R$\sim$24), wide ($\sim$0.5
deg$^2$) multiband (BVRI Vega system) images of the Coma cluster obtained 
with the
CFH12K camera at the CFHT. We have derived LFs and CMRs in twenty
10$\times$10~arcmin$^2$ regions and in larger regions.}
{In all photometric bands, we found steeply rising LFs in the
north-northeast half of the cluster (due to early type galaxies at bright
magnitudes and due to late type galaxies at the faint end), 
and much flatter LFs in the
south-southwest region. Although the fine behaviour of the CMR RS is
different in these two regions, a good agreement is found in general
between the RS computed for faint and for bright galaxies.}
{All these results can be interpreted consistently in the framework of
the building up process proposed by Adami et al. (2005b). The Northern
Coma area is a relatively quiescent region while the southern area
experiences several infalls.}


\maketitle

\section{Introduction}\label{sec:intro}

Hierarchical building of clusters is a key ingredient of cosmological
models, since clusters are believed to be at the nodes of 
cosmological filaments and to grow by accreting matter (gas, galaxies,
groups etc.) from filaments (e.g.  Lanzoni et al. 2005).  Such
continuous infall has major consequences on the shape and dynamics of
clusters, as well as on the physics of cluster galaxies (e.g. Sarazin
1986) as revealed by the cluster galaxy
mass function (usually traced by the luminosity function, LF
hereafter) and by the cluster early type galaxy Red Sequence (RS
hereafter). 

The influence of infall on the LF is relatively
well understood for bright galaxies in nearby to intermediate redshift
clusters (e.g. Hansen et al.  2005), but only little is known for low
mass galaxies because they are difficult to observe. It is also
important to study galaxies at the faint end of the RS because the
Color Magnitude Relation (CMR hereafter), which traces their star
formation history, is directly related to their metal abundances and
is poorly known. These faint galaxies are of major interest as their
evolutionary paths are sometimes different from those of bright
galaxies: they are very sensitive to environmental effects
(e.g. fusion with bright galaxies or tidal disruptions); they also
keep their dynamical memory longer (e.g.  Sarazin 1986) and their spatial
distribution is different from that of bright galaxies
(e.g. Biviano et al. 1996, but see also Edwards et al. 2002 for a different
result).

To study the properties of galaxy clusters through their LF and CMR
requires to sample the whole cluster galaxy populations from the
bright cD to the faintest dwarfs (M$_R \sim -10$). Very demanding
observations are needed both in terms of photometric depth and size of
field, and were not possible until the arrival of wide-field
facilities such as the CFH12K on the CFHT. With such a 4m-class
telescope, access to the faintest objects is still limited to nearby
clusters. In this respect, the Coma cluster is possibly the best
target: bright objects are well studied in all wavelengths
(e.g. Biviano 1998) and its low redshift allows to go deep enough
towards the faint end of the galaxy population.

LFs and CMRs have been recently computed for bright to intermediate
magnitudes in Coma (e.g. Lobo et al. 1997, Terlevich et al. 2001,
Andreon $\&$ Cuillandre 2002, Beijersbergen et al. 2002,
Iglesias-P\'aramo et al. 2003) in large fields. Deeper studies in
rather limited areas also exist (Bernstein et al. 1995 or Trentham
1998). All these surveys generally have a wavelength coverage too
limited to provide strong constraints on the Coma cluster building
history. For the first time, by using the CFH12K camera at the CFHT,
we obtained a set of deep (R$\sim$24), wide ($\sim$0.5 deg$^2$) 
multiband (BVRI) images of the Coma cluster.

The key point to compute a LF or a CMR is to select objects in the
cluster and then galaxies, as compared to intergalactic Globular
Clusters and galactic stars. For the
bright part of our sample, we had enough spectroscopic redshifts to
discriminate Coma members.  As it is prohibitive to obtain spectra for
the faint part, we used statistical comparisons with empty fields
(containing no rich nearby clusters). This is a classical method which
requires very homogeneous observations of the cluster and of the empty
fields.  The empty fields we used were parts of two of the VVDS CFH12K
imaging survey fields (McCracken et al. 2003: MC03 hereafter)
completely re-analyzed in order to have catalogs homogeneous with the
Coma data. This allows to treat properly possible pitfalls such
as diffuse light in the cluster, galactic extinction, differences in
the star counts, large scale structure of the background and crowding
effects.  Other problems which
could affect the background counts along the Coma line of sight are
the presence of dust in the cluster or the gravitational effects on
background galaxies but they have been shown to be negligible in Coma
(Bernstein et al. 1995).

In Section 2, we review the data and catalogs and we describe the
homogenization process of these catalogs in Section~3. In Sections 4
and 5, we compute the LFs for bright and faint galaxies. Section 6 is
dedicated to the nature of the faintest Coma cluster objects. In
Section 7, we describe the LF characteristics. Section 8 describes the
RS in the Coma CMR. We discuss in Sections 9 and 10 the inferences of
the Coma LF and CMR RS on the cluster building up scenario. Finally,
Section 11 is the conclusion.

We assume a distance to Coma of 95 Mpc,
H$_0$ = 75 km s$^{-1}$ Mpc$^{-1}$. The distance modulus is 34.89,
and therefore the scale is 0.46 kpc arcsec$^{-1}$. We give magnitudes in the
Vega system except if specified otherwise.

\section{Raw data and catalogs}

\subsection{Data}

We use here the Coma data described in Adami et al. (2006a: A06a
hereafter) and already used in Adami et al. (2005a) and Adami et
al. (2006b). These data were obtained at CFHT with the CFH12K camera
using Johnson-like filters in four bands (BVRI), and reach R$\sim$24
in an area of $\sim$0.5 deg$^2$. We refer the reader to A06a for more
details.

In order to have a homogeneous set of empty fields, we used the VVDS
CFH12K imaging fields F02 and F10 (other VVDS CFH12K imaging fields 
do not have the same wavelength coverage). These two fields were observed with
the same B, V, R and I filters (see MC03 for a full description of the
F02 field) and have been checked to be free from rich nearby clusters
(using VVDS preliminary spectroscopic data). We only used areas where 
the four bands were simultaneously available. As described later, we 
completely re-extracted from these images the catalogs with
the same parameters in the most homogeneous possible way compared to Coma data.

A crucial point is that a non negligible uncertainty source in the LF
computation comes from the cosmic variance between the Coma line of
sight and the comparison fields and inside the comparison fields
themselves. As shown later, the total area of our comparison fields is
almost equivalent to a 45$\times$45 Mpc$^2$ area at z$\sim$0.7 (the
mean F02 sample redshift, Le~F\`evre et al. 2005). This is close to
the mean size of large scale structures (e.g. Hoyle $\&$ Vogeley
2004).

\subsection{Object extraction}

Besides having been observed with the same telescope, instrument and
filters, all the data used here were obtained in similar seeing
conditions (ranging from 0.8 to 1.07~arcsec, see A06a and MC03).
However the catalogs presented in MC03 have been extracted using much
less stringent SExtractor (Bertin $\&$ Arnouts 1996) parameters than
in our Coma cluster catalog, giving a quite different completeness.
So we decided to re-extract from the F02 and F10 images our own 
catalogs in the Vega system with
exactly the same SExtractor parameters as in A06a. These are mainly a
detection threshold of 2$\sigma$ and a minimum number of contiguous
pixels above the detection threshold of 9.

\begin{figure}
\centering
\mbox{\psfig{figure=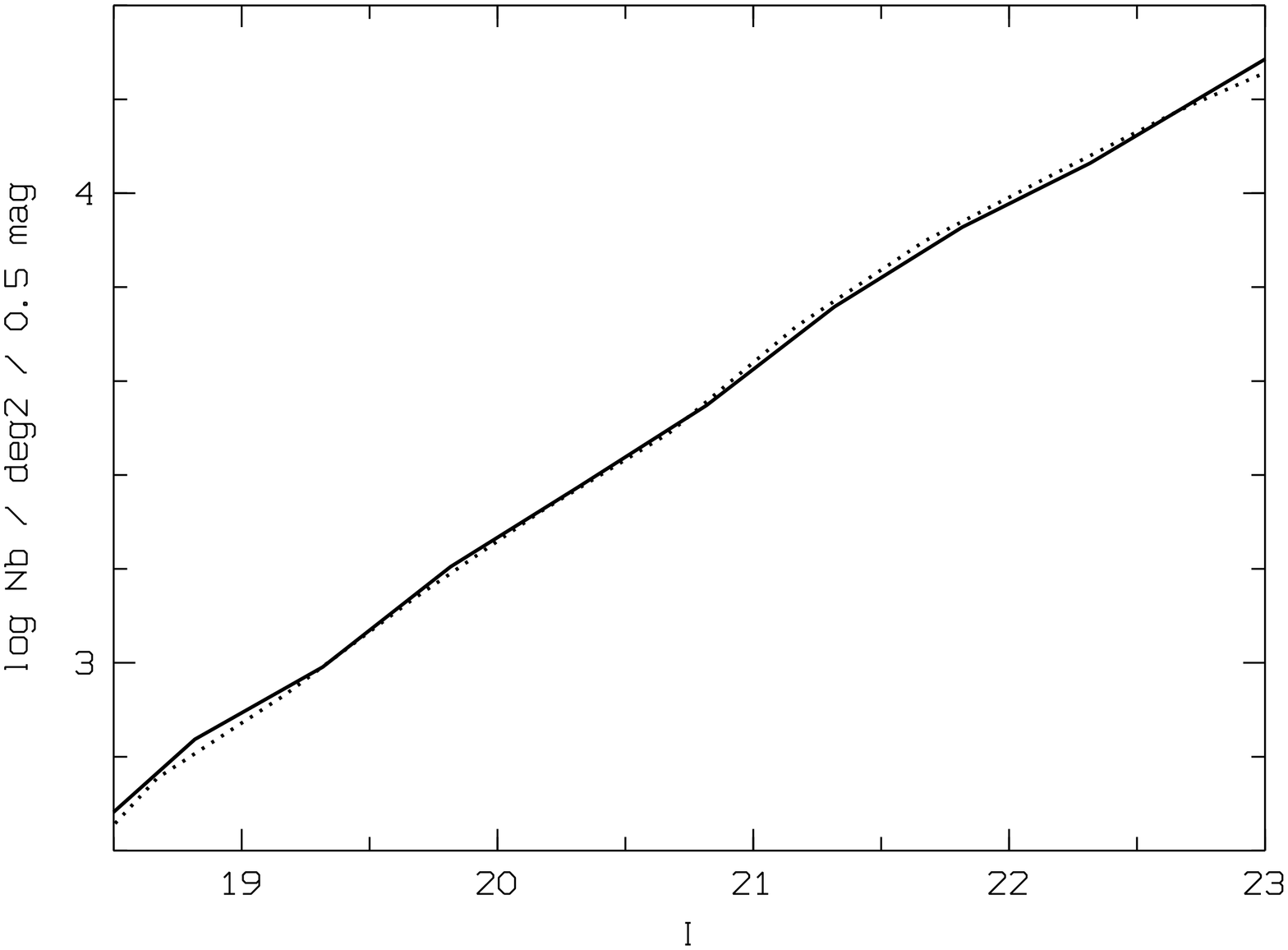,width=8cm,angle=270}}
\caption[]{Differential galaxy counts in the F02 field. Continuous
line: original MC03 catalog. Dotted line: our own
extraction. The I magnitude range is limited to our mean completeness
level: I=23. }
\label{fig:cra}
\end{figure}

We check that our counts, up to our completeness limit, are in good
agreement with those of MC03 for the F02 field in the I
band. Corrected by a shift of 0.433 to be translated into the Vega
system, we found the MC03 I band magnitudes to be 0.08 magnitude
brighter than ours (matched to the synthetic star locus, cf 3.1).
This does not mean that the MC03 I band magnitudes are wrong by this
amount but that there is a shift of 0.08 magnitude between the MC03 I band
magnitudes and the spectral templates we used (which are probably not
perfect). Applying this shift and comparing the two
galaxy counts, we see on Fig.~\ref{fig:cra} that the agreement is very
good.

Finally, since the Coma and the two empty fields have different
galactic extinctions, we make a correction using the Schlegel et
al. (1998) maps.  For the F02 field, we applied a mean magnitude
correction on all galaxies because the extinction was only weakly
varying across the field with an uncertainty of less than $\sim$0.01
in the four bands.  For the F10 field, we had to correct locally each
galaxy using an extinction map with a spatial resolution of 30
arcsecs.

\subsection{Star-galaxy separation}

Since the three fields are at different galactic latitudes, we need to
remove stars in a homogeneous way in all the fields. We therefore
performed the same star-galaxy separation on the three fields,
distinct from the method used by MC03. Our own separation was based on
the comparison of total and central I band magnitudes.  We started by
dividing each of the three fields (Coma, F02 and F10) into subfields
of $\sim$0.3 deg$^2$ (two subfields for Coma, four for F02 and six for
F10) in order to deal with possible PSF variations across the fields
that could have an effect on the star-galaxy separation. We show in
Fig.~\ref{fig:sgF02} an example of the star-galaxy separation in one
of the F02 subfields and we refer the reader to A06a for a full
description of the Coma star-galaxy separation and of the general
method. We did not use one of the F10 subfields which did not allow an
efficient star-galaxy separation because of various star-sequences
appearing in the total vs. central I band magnitude plots, probably
because of a strongly varying PSF.

Our star-galaxy separation could probably misclassify extremely
compact galaxies in the magnitude range where our separation is efficient 
(I brighter than 21).  However, we showed in A06a
(Fig. 15) that our star counts match very well star counts from the
Besan\c con model of our Galaxy (Gazelle et al. 1995).  The difference
between the two counts is 4$\pm$11$\%$ in each magnitude bin. 
The number of extremely
compact bright galaxies misclassified is therefore not a serious
concern.

\begin{figure} \centering
\caption[]{Star-galaxy separation based on the central I surface 
brightness (in mag/arcsec$^2$) versus total I magnitude. The
thick (red) solid line indicates the separation we applied (in the
magnitude range delimited by the two thin (green) vertical continuous
lines) between stars and galaxies for one of the F02 subfields. }
\label{fig:sgF02}
\end{figure}

\subsection{Saturated objects and field masking}

As shown in A06a, a significant number of galaxies brighter than
I$\sim$18 are saturated in the Coma field. The F02 and F10 fields are
also subject to the same saturation effects and we chose to use the
statistical comparison between the Coma and empty fields only for
magnitudes fainter than this value. For brighter magnitudes, the LF
will be derived with another method described later.

Similarly, very bright objects have extended halos that lower
significantly object detection and increase the magnitude error in the
concerned regions. We therefore chose to exclude from all fields the
areas located at less than twice the radius of objects brighter than
I=18. Some regions showing low signal to noise or diffraction spikes
were also excluded manually from the sample.

Fig.~\ref{fig:pos} shows the position of all the galaxies in the Coma
and F02 fields, the rejected regions and the splitting into subfields
chosen to perform the analysis.

\begin{figure}
\centering
\caption[]{Top: map of detected objects along the Coma cluster line of
sight, showing rejected zones in white. The solid line delineates the
two Coma subfields. Bottom: same for the F02 field and subfields.
$\alpha$ and $\delta$ are given in decimal degrees, $\alpha$
increasing to the right. We clearly see in the upper figure the
shallower depth of the Coma north data.}
\label{fig:pos}
\end{figure}

\subsection{Area coverage}

We finally obtained a coverage of $\sim$1.2 deg$^2$, $\sim$1.6 deg$^2$
and $\sim$0.5 deg$^2$ for the F02, F10 and Coma fields
respectively. We note that a small part of the Coma field (about 200
arcmin$^2$ at the cluster south west) was only available in the R and
I bands. We summarize in Table~\ref{tab:area} the area coverages in
each band for each of the fields and subfields.

\begin{table}
\caption{Area coverage for each of the comparison subfields. }
\begin{tabular}{cll}
\hline
Subfield & Useful area (deg$^2$) & Masked percentage \\
\hline
F02 subfield 1 & 0.3087 & 17$\%$ \\
F02 subfield 2 & 0.3153 & 15$\%$ \\
F02 subfield 3 & 0.2958 & 21$\%$ \\
F02 subfield 4 & 0.3049 & 18$\%$ \\
F10 subfield 1 & 0.2190 & 21$\%$ \\
F10 subfield 2 & 0.5828 & 37$\%$ \\
F10 subfield 3 & 0.3016 & 30$\%$ \\
F10 subfield 4 & 0.3715 & 10$\%$ \\
F10 subfield 5 & 0.1340 & ~4$\%$ \\
\hline
\end{tabular}
\label{tab:area}
\end{table}

The total physical area of our comparison fields is about 45$\times$45
Mpc at z$\sim$0.7. This is of the same order as the periodicity
of the Universe large scale structures (e.g. Hoyle
$\&$ Vogeley 2004) and ensures that we will not subtract field counts
only coming from a peculiar under or over-dense field area.

Table~\ref{tab:compar} presents a comparison between this survey and
previously available data from the literature in different bands, in
terms of photometric depth and surveyed area (both in Coma and in the
comparison fields).  This clearly shows that among all studies, our
survey is the best compromise between depth, coverage and robustness
in the field comparison.

\begin{table}
\caption{Comparison with recent Coma cluster data from the
literature. We quote in this table our point source 90$\%$
completeness levels. The completeness levels of literature data are
those quoted in the corresponding papers. We also give the Coma
cluster areas covered, together with the surface of the comparison
fields in each case.}
\begin{tabular}{cllll}
\hline
Filter & Observed & Limiting  & Control & Reference \\
       & surface  & magnitude & field   & \\
       & (deg$^2$)&           & (deg$^2$) &  \\
\hline
B      & 0.45     & 24.75 & 2.8  & present survey \\
       & 0.20     & 22.5 & 0.17  & AC02 \\
       & 5.2      & 21.7 & 0.556 & Bj02 \\
       & 0.187    & 24.0 & 0.037 & T98 \\
V      & 0.45     & 24.0 & 2.8 & present survey \\
       & 0.29     & 23.75 & 0.18 & AC02 \\
       & 0.417    & 22.0 & 0.4   & L97 \\
R      & 0.50     & 24.0 & 2.8 & present survey \\
       & 0.29     & 23.25 & 0.18 & AC02 \\
       & 5.2      & 21.7 & 0.278 & Bj02 \\
       & 0.0145   & 25.5 & 0.075 & B95 \\
       & 0.97     & 20.5 & SDSS  & IP03 \\
       & 0.187    & 23.5 & 0.052 & T98 \\
I      & 0.50     & 23.25 & 2.8 & present survey \\
\hline
\end{tabular}
\label{tab:compar}
Notes to Table~\ref{tab:compar}: AC02 = Andreon \& Cuillandre (2002),
Bj02 = Beijersbergen et al. (2002), B95 = Bernstein et al. (1995),
IP03 = Iglesias-P\'aramo et al. (2003), L97 = Lobo et al. (1997),
T98 = Trentham (1998).
\end{table}

\section{Catalog homogenization}

\subsection{Magnitudes}

Computing LFs based on statistical empty field comparisons requires to
subtract the empty field galaxy counts from the Coma cluster galaxy
counts. As these counts have very steep slopes at faint magnitudes,
even small systematic magnitude uncertainties between the fields can
induce significant errors in the final results (e.g. Lobo et
al. 1997). We therefore require relative uncertainties between the
different fields smaller than 0.05 magnitude in each photometric band;
this upper limit leads to differences of the order of 40$\%$ in our LF that
are already prohibitive.

Using the field and Coma catalogs as they are does not generally allow
to reach such a precision because of possible small zero point
variations or PSF inhomogeneities. Even relatively modest seeing
variations of $\sim 10 \%$ can induce differences of several tenths of
magnitudes for faint objects (e.g. Savine 2002), too large for our
requirements. We therefore decided to rescale all bands in order to
match the spectrophotometric star template loci in color-color
plots. This does not mean that we get the true magnitudes, but that
magnitudes in the different fields will be homogenized with a
precision of the order of a few 0.01 magnitudes.

In each of the different subfields, we selected stars (as defined in
A06a) and compared in all the possible color-color plots (using B, V,
R and I bands) the observed and template star loci. Templates were
taken from the library of Pickles (1998). Results after the rescaling
are shown in Fig.~\ref{fig:shift} for one of the subsamples of the F02
field and for one color combination. The values of the individual
magnitude shifts applied in each subfield and for each band are
0.01$\pm$0.04 magnitude, smaller than the 0.05 magnitude limit. Even
if they are low, showing that our fields were already quite
homogeneous, the effects of these corrections are not negligible on
the LF errors.

We finally note that the corrections applied to galaxies were computed
from star loci. Even though the surface brightness profiles of bright
galaxies are different from those of stars, synthetic color-color
sequences for stars can still be used advantageously as secondary
photometric standards. Indeed, our study is mainly focused on faint
galaxies which are at least partially seeing dominated, as stars are.

\begin{figure}
\centering
\caption[]{Color-color plots for one of the F02 subfields after
applying the small magnitude shifts described in Section~3.1. Small
dots are the stars extracted from the F02 observations and circles
are the star synthethic models. }
\label{fig:shift}
\end{figure}

\subsection{Surface brightness limitations}

As shown in Fig.~\ref{fig:surfmag}, the F02 and F10 data are deeper
than our Coma data in terms of total magnitude and surface
brightness. Therefore, in order to avoid oversubstracting background
objects to the Coma counts, we limited both the Coma and the F02 and
F10 catalogs to a given surface brightness (I=25.2, R=26.0, V=26.4 and
B=26.8 mag arcsec$^{-2}$). These values were chosen in order to
maximize the differences between the Coma and empty field counts as a
function of magnitude. Fig.~\ref{fig:surfmag} illustrates this case in
the I band.

\begin{figure}
\centering
\caption[]{Mean I surface brightness (in mag/arcsec$^2$) versus total
I magnitude. Crosses are objects from the F02 field and light (green)
dots are from the Coma cluster line of sight. The horizontal line
indicates the surface brightness cut we applied. The blue triangle
indicates the place where low surface brightness objects are in excess
in the cluster compared to the field.}
\label{fig:surfmag}
\end{figure}

It is also interesting to note that an excess of low surface
brightness galaxies appears in the Coma field, at I lower than 21.5,
a peculiar class of objects already detected in Adami et al. (2006b).

\subsection{Total magnitude completeness limits}

Completeness is not straightforward to estimate and is a complex
function of galaxy characteristics and of detection parameters. For
example, relatively compact galaxies have a concentrated brightness
profile and are easier to detect than low surface brightness galaxies.

Using simulations, we defined in A06a two completeness limits in each
of the observed bands : the 90$\%$ point-source completeness and the
90$\%$ extreme low surface brightness galaxy completeness. These two
estimates give the interval where the completeness level should lie
for a given class of galaxies. For example, we show that a 90$\%$
completeness level is reached between R=20.75 and R=24 in the R band 
(north data) and between I=20.25 and I=23.25 in the I band 
(south data) (see A06a).

A second way to estimate the mean completeness over all classes is to
compare our catalog with deeper catalogs.  Using the Bernstein et
al. (1995) R band catalog (hereafter B95), we found a mean R band
90$\%$ completeness level of R$\sim$23.5 (cf. A06a).

We present here a third method close to that described in Andreon $\&$
Cuillandre (2002) and using only the catalog data. This method will
allow us to give an estimate of the mean completeness level in all 
bands, resulting from the mean object profile over the whole cluster
galaxy population. We proceed as follows:

- First we use the total magnitude range where our data are deep
enough to detect all galaxies, whatever their surface brightness.  We
then consider the upper central surface brightness envelope of the
galaxy population as a function of total magnitude, which is only
related to the physical constraints between maximum surface brightness
and total magnitude. This limit is represented by inclined lines in
Fig.~\ref{fig:surfmag2}. 

- Second, we define our surface brightness detection limit (horizontal
lines in Fig.~\ref{fig:surfmag2}) which is directly linked to our
survey characteristics.

The intersection between these two lines gives a conservative estimate
of the completeness (method illustrated in
Fig.~\ref{fig:surfmag2}). This method gives completeness levels of
I$\sim$23 and R$\sim$23.5. The R value is very similar to that given
by the comparison with the B95 catalog. The R and I values are
brighter by 0.5 magnitude than computed with point source simulations
(A06a). We therefore decided to assume a mean completeness 0.5
magnitude brighter than the point source 90$\%$ completeness levels 
(given in A06a and that are different between north and south data)
whatever the band, namely, B=24.25, V=23.5, R=23.5 and I=23 for the
mean Coma cluster line of sight galaxy population. 

It can be noted that the southern field is deeper than the 
northern field in most of the bands (see A06a), as extensively discussed in
A06a.  We stress, however, that in the present paper we use in each
magnitude band the brightest magnitude limit of the north AND south
fields to define the completeness levels. This ensures that no
artificial effects will pollute our results: in the magnitude limits
we fix, our data are complete whatever the location in the cluster.

\begin{figure}
\centering
\caption[]{Central surface brightness (in mag/arcsec$^2$) versus total
magnitude. The thick inclined line gives the intrinsic maximum
central surface brightness for a given total magnitude. The 
horizontal line gives the maximum central surface brightness
detectable in our Coma data. The intersection between the two lines
gives an estimate of the mean completeness level. Top: R band, bottom:
I band.}
\label{fig:surfmag2}
\end{figure}

\section{Bright galaxy luminosity functions}

We will now describe how we can compute the bright galaxy LF and how
it compares to numerical simulations. Computing a LF using redshifts
is the most accurate method and must be used as much as possible, but
of course it is difficult to achieve redshift completeness down to
faint magnitudes.

\subsection{Spectroscopic luminosity functions}

As described in A06a or Adami et al. (2005b), we have compiled the
largest sample of spectroscopic redshifts presently available
along the Coma cluster line of sight ($\sim$500 redshifts
in the area covered by our CFH12K data). 

However we are not 100$\%$ complete in spectroscopy even at bright
magnitudes. We therefore applied statistical arguments by computing
for a given magnitude the percentage of galaxies with a redshift
inside Coma, and applying this percentage to the total number of
galaxies (with and without a redshift) inside the bin. We also limited
our sample to a $\sim$50$\%$ completeness level. Translated in
magnitudes, this gives the following faint ends for the spectroscopic
LFs: I=18.25, R=18.75, V=18.75 and B=19.25.

Uncertainties on these LFs due to incompleteness are then simple to
compute, since the minimal number of galaxies inside Coma is the number of
galaxies with a known redshift inside Coma and the maximum number is
the total number of galaxies inside the magnitude bin. This range
gives an estimate of the $total$ error (100$\%$), close to a 3$\sigma$ error.

\begin{figure}
\centering
\mbox{\psfig{figure=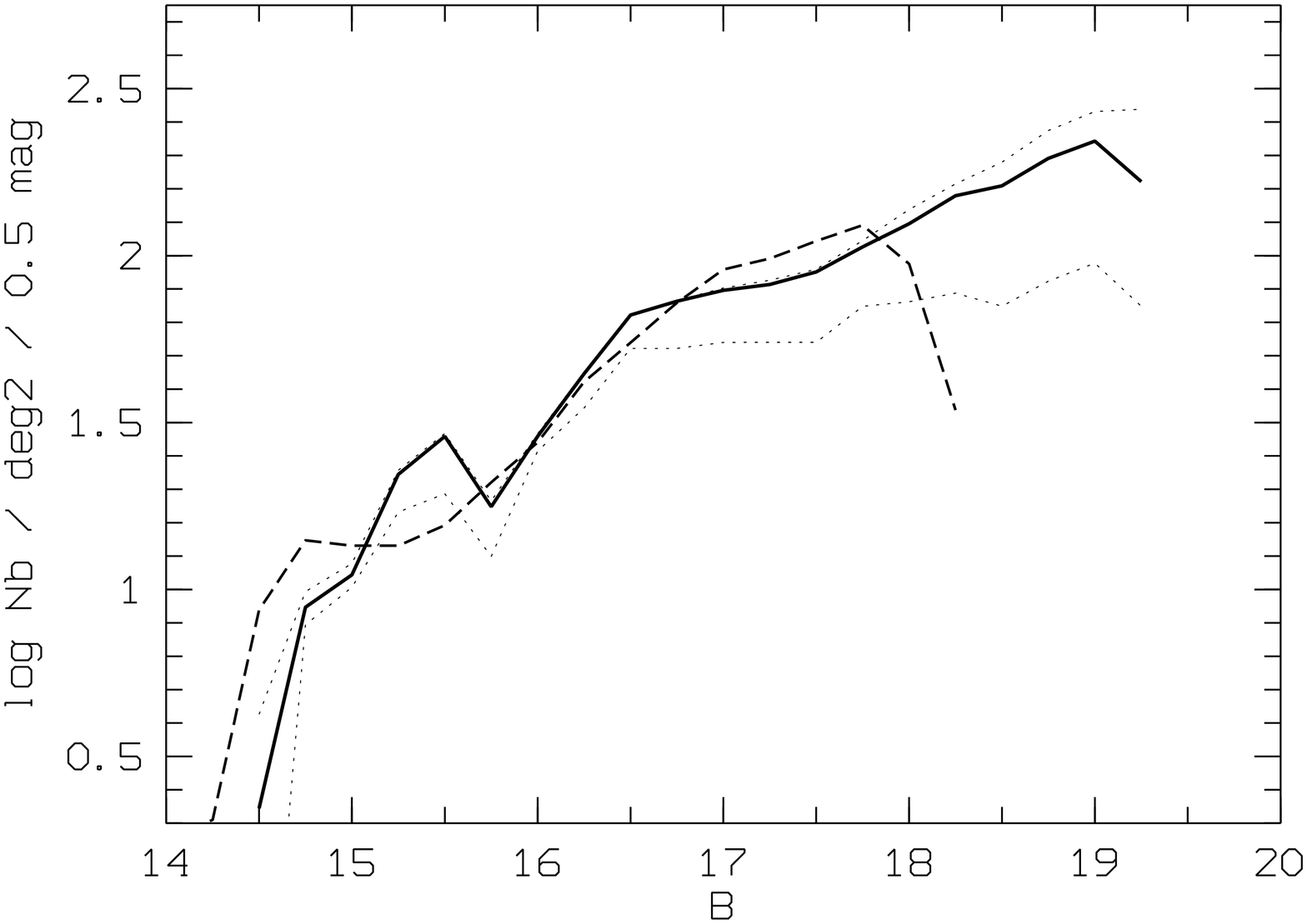,width=8cm,angle=270}}
\mbox{\psfig{figure=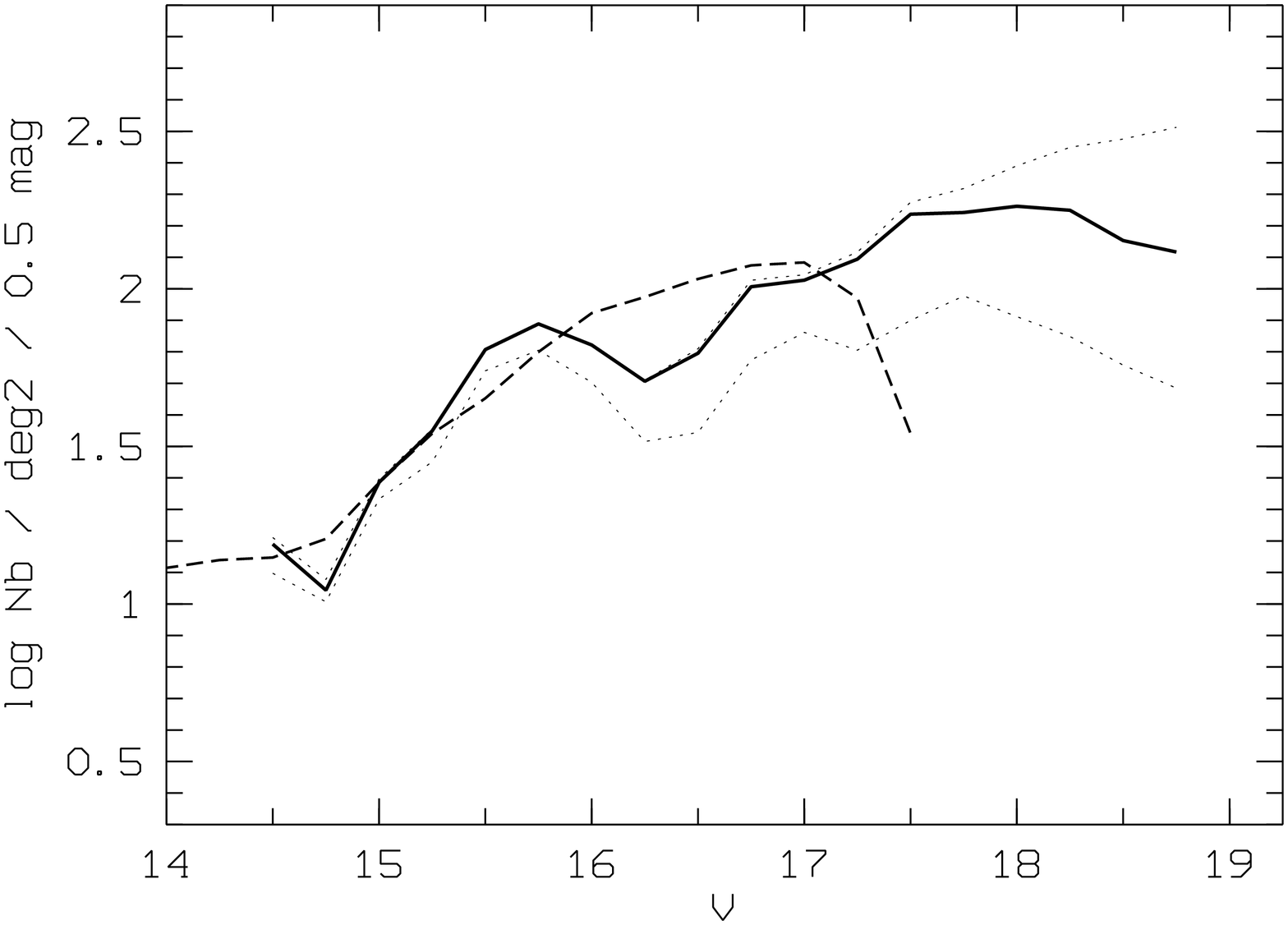,width=8cm,angle=270}}
\caption[]{Coma cluster galaxy LFs computed with spectroscopic
redshifts (continuous line) along with the $total$ error envelope
(dotted lines). From top to bottom: B and V. Simulated LFs are 
shown as a dashed line.}
\label{fig:fdlbright}
\end{figure}

Bright LFs are shown in Fig.~\ref{fig:fdlbright} for the B and V bands
together with their 3$\sigma$ uncertainties which mainly reflect the
redshift catalog incompleteness. LFs are computed in a 0.5 magnitude
running window with a step of 0.25 magnitude.

\subsection{Comparison with simulations}

In order to compare our bright LF with modeled clusters we used the
simulations described in Lanzoni et al. (2005). Their mass resolution
is 10$^{9}$ h$^{-1}$ M$_\odot$, corresponding to an absolute B
magnitude of $\sim -17$ (B$\sim$18 at the Coma cluster redshift). This
is not sufficient to produce faint galaxies, but a comparison with
real data will allow to validate the simulation process and check that
observations and simulations are compatible.

We extracted a LF from the simulations inside a circle of 0.3 virial
radius, the same physical area than our Coma cluster data. We used a
r$_{200}$ virial radius of 1.5 h$^{-1}$ Mpc (Geller et al. 1999) for
the Coma cluster.  We renormalized the number of galaxies by the ratio
between the Coma cluster mass in the considered area and the total
mass of the systems simulated by Lanzoni et al. (2005), based on the
15 simulated clusters with masses ranging between 10$^{14}$ and
10$^{15}$ M$_\odot$.

The result of the comparison is shown in Fig.~\ref{fig:fdlbright} in
the B and V bands. The agreement appears very good, except for the
facts that:

- the LF appears smoother in the simulated clusters than in
Coma

- the simulated LF does not follow the observations above  V$\sim$17 and
B$\sim$18 because of the lack of mass resolution.

The dips seen in the Coma cluster LFs (significant given the
3$\sigma$ uncertainties) are probably due to peculiar processes in the
Coma cluster, not reproduced by generic cosmological simulations and
that we will discuss later.

\section{Faint galaxy luminosity functions}

To select the objects in Coma beyond the available spectroscopic
data, we use two methods described in this section along with the
resulting uncertainty estimates.

\subsection{Faint luminosity function using simple statistical subtractions}

In this first approach, the number of galaxies is expressed along the
Coma cluster line of sight as a function of magnitude, whatever their
colors:

N$_{C~los}$(mag) = N$_{C}$(mag) + N$_{efield~C}$(mag)

\noindent
with N$_{C~los}$ the number of objects along the Coma line of sight,
N$_{C}$ the number of objects inside the Coma cluster and
N$_{efield~C}$ the number of background and foreground objects at the
Coma position (empty field).

From the F02 and F10 fields, we obtained an estimate of
N$_{efield}$(mag) in the field but at positions different from the
Coma cluster. If our comparison fields are large enough to smooth the
field-to-field variance, then N$_{efield}$(mag) - N$_{efield~C}$(mag)
should have a null mean and we get:

N$_{C}$(mag) = N$_{C~los}$(mag) - N$_{efield}$(mag)

\noindent
with an uncertainty which will be estimated in the following.

LFs are computed in a 0.5 magnitude running window with a step of
0.25 magnitude.

\subsection{Faint luminosity function using color-dependent statistical
subtractions}

This time, we use the same method, considering the galaxy
counts not only as a function of magnitude, but also as a function of
colors, to allow for example selections on colors and estimates of LFs
for different classes of galaxies.

The estimate of the Coma cluster counts then becomes:

 N$_{C}$(mag,col$_i$) =  N$_{C~los}$(mag,col$_i$) -
  N$_{efield}$(mag,col$_i$) 

\noindent
with col$_i$ being all colors available.

N$_{C}$(mag) is then the sum of N$_{C}$(mag,color$_i$) over
all color$_i$ in the galaxy sample. 

A good way to understand this type of method is to consider only one
color. In this case, the method basically consists in estimating the
difference between the Coma cluster line of sight and the empty field
F02 and F10 CMRs. As an illustration, we will describe in the
following the computation of the B-R/R CMR and the detection of a RS
inside the Coma cluster down to very faint magnitudes.

\subsection{Luminosity function uncertainties}

Uncertainties on the counts are of several origins.

The first uncertainty is the poissonian error simply related to the
limited number of galaxies in each magnitude bin in the empty fields
or in the Coma field.

Other sources of errors are related to the magnitude accuracy and to
the distribution of galaxies in the different fields.

Magnitude uncertainties lead to errors on galaxy counts that propagate
into LF uncertainties when computing the difference between the Coma line
of sight and the empty field counts. These uncertainties can be
estimated as a function of magnitude with Monte-Carlo simulations by
generating catalogs of objects with magnitudes gaussianly distributed
according to the data statistics given in A06a, and computing 100
LFs. The whole sample of generated LFs therefore allows to estimate
the uncertainty as a function of magnitude.

Another source of uncertainties on the background counts and LFs, is
the field-to-field variance and a way to deal with this effect is to
use many comparison fields and get the field-to-field variance from
the fluctuations on the measured counts.  This method was used for
example by Bernstein et al. (1995), but as we do not have enough
independent comparison fields (only two) we use a method applicable on
a single field and described by Huang et al. (1997).  Andreon $\&$
Cuillandre (2002) have underlined the fact that relatively strong
variations of the luminosity and correlation functions induce
variations of only a few percent on the errors induced by the cosmic
variance. However, we have decided to estimate this error. Following
Huang et al. (1997), we computed the correlation function in the F02
field using the VVDS redshifts with the selection criteria used for
our photometric samples. We then assumed the correlation function of
the F10 and of the Coma line of sight to be the same. Regarding LFs,
we used the Ilbert et al.  (2005) study to estimate a mean value of
the Schechter parametrisation within our selection criteria.  The
parameters we used are given in Table~\ref{tab:cv}.

\begin{table}
\caption{Parameters used for empty field luminosity and correlation functions.}
\begin{tabular}{lllll}
\hline
Filter & B & V & R & I \\
\hline
Schechter $\alpha$ &  -1.23   & -1.39  &  -1.44  & -1.45  \\
Schechter Vega M$^*$ &  -21.0   & -22.0  &  -22.7  & -23.3  \\
Power law $\gamma$ &  1.596   & 1.548  &  1.612  & 1.628  \\
Power law r$_0$ &  2.485   & 2.545  &  2.575  & 2.725  \\
\hline
\end{tabular}
\label{tab:cv}
\end{table}

Finally, all these errors were quadratically added to obtain the error
budget for the faint parts of the LF.

\begin{figure}
\centering
\mbox{\psfig{figure=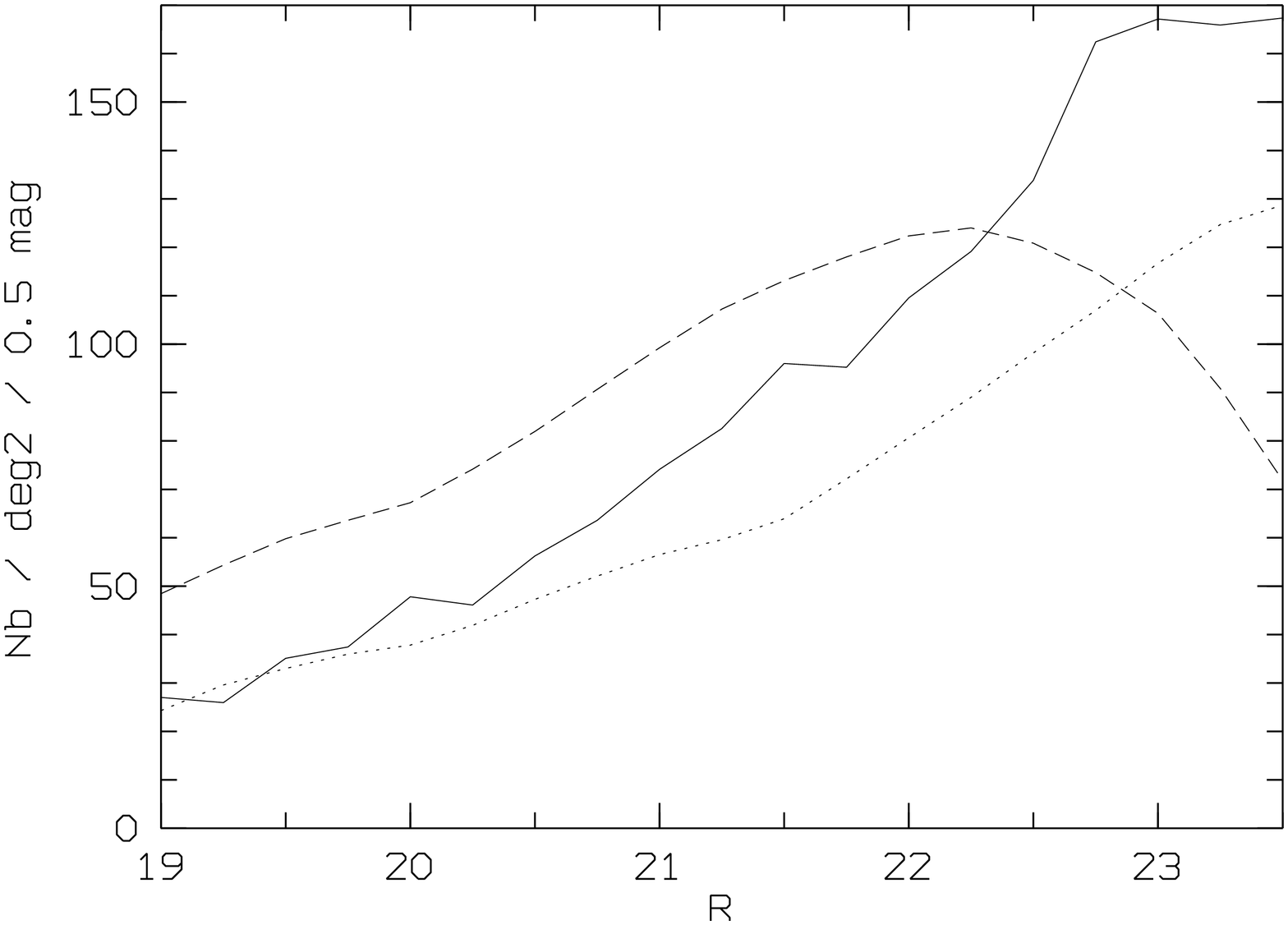,width=9cm,angle=270}}
\caption[]{Three uncertainty sources for the zone of Coma in the R
band: dashed line: error from the field-to-field variance, solid line:
error from magnitude uncertainties, dotted line: poisson error.}
\label{fig:erreurs}
\end{figure}

Fig.~\ref{fig:erreurs} shows the three contributions to the LF
uncertainty in the Coma area in the R band. At bright magnitudes, the
field-to-field variance error dominates, while the error due to the
uncertainty on magnitudes is the largest one at faint magnitudes.

Given the relatively low level of uncertainty due to
field-to-field variance, the counts of galaxies between the F02 and
F10 fields should be very similar. Between R=19 and 23.5, the
difference between the numbers of galaxies in the F02 and F10 fields
in 0.25 mag bins is indeed very small: $-71 \pm$120
galaxies/deg$^2$/0.25mag; this represents about 2$\%$ of the total
number of galaxies inside the considered bins.

\section{Expected contamination by Globular Clusters at the faint end
of the LF }

The faintest objects found to be in the Coma cluster by the
statistical subtraction method can be either galaxies or globular
clusters (GC hereafter). Using a rather small area, Bernstein et
al. (1995) have shown that the GC contribution becomes significant for
R fainter than 23.5. However, Mar\'\i n-Franch $\&$ Aparicio
(2003) reach apparently contradictory conclusions.  Therefore, in
order to estimate the relative fractions of these objects in our data
we used the description of the GC population around NGC~4406 given by
Rhode $\&$ Zepf (2004). This galaxy (also known as M86) is an
elliptical cluster galaxy attached to the Virgo cluster. Its GC
population is therefore typical of the GCs we could detect in Coma.

According to Rhode $\&$ Zepf (2004) GCs extend to less than twice the
typical galaxy optical radius from the NGC~4406 center, and given our
masking conventions near bright galaxies, the GCs still attached to
galaxies brighter than I=18 are not counted in our LFs. The only GCs
counted are those attached to fainter galaxies (a probably minor
contribution) and those expelled from their parent galaxies that we 
will discuss now.

The brightest GC in NGC~4406 has a magnitude of V$\sim$20, which
translated to the Coma cluster distance (assuming a Virgo distance of
16.7 Mpc) gives a V magnitude of $\sim$23.75. This means that the
population of single GCs detected in our data is probably negligible at
V brighter than 23.75 and we will only consider the fainter magnitude
bins as potentially polluted by GCs.

To estimate the number of GCs in the next $\sim$half magnitude bin, we
used the color-color plot of Rhode $\&$ Zepf (2004), which gives
the location of their candidate GCs in the B-V versus V-R plane. Their
B and V filters are very close to our own filters (Johnson-like
filters) while their R filter is close to a Cousins filter and we
correct their R values by $-0.1$ mag assuming that GCs have
elliptical-like colors (Fukugita et al. 1995). With the same
assumption, we also shifted the B-V and V-R colors from the Virgo
distance to the Coma distance.

We then computed the B-V versus V-R map of objects statistically in
the Coma cluster (obtained by subtracting the empty field B-V versus
V-R map from the Coma line of sight B-V versus V-R map, after
renormalizing them by their respective areas), 
in which we approximately defined the GC location from Rhode
$\&$ Zepf (2004). This allowed us to compute the percentages of
objects in these maps located at the same place as GCs. These
percentages, given in Table~\ref{tab:GCs}, are only upper values of
the number of globular clusters as regular faint Coma cluster galaxies
can have similar colors.

\begin{table}
\caption{Percentage of total number of objects with colors similar to
GCs for the Coma subfields (see Figs.~\ref{fig:fdl1}, ~\ref{fig:fdl2},
~\ref{fig:fdl3} and ~\ref{fig:fdl4}). Typical 1$\sigma$ errors on
the total number counts are also given.}
\begin{tabular}{llllllll}
\hline
\hline
Subfield & 4 & 5 & 8 & 9 & 10 & 13 & 14 \\
\hline
$\%$ & 25$\%$ & 37$\%$ & 66$\%$ & 37$\%$ & 35$\%$ & 8$\%$ & 30$\%$ \\
\hline
error & 21$\%$ & 20$\%$ & 21$\%$ & 21$\%$ & 20$\%$ & 18$\%$ & 22$\%$ \\
\hline
\hline
Subfield & 15 & 18 & 19 & 20 \\
\hline
$\%$ & 52$\%$ & 14$\%$ & 32$\%$ & 36$\%$ \\
\hline
error & 20$\%$ & 20$\%$ & 20$\%$ & 21$\%$\\
\hline
\hline
\end{tabular}
\label{tab:GCs}
\end{table}

The result is a complex interplay of object colors and of course,
these percentages are only valid in the magnitude ranges in which the
completeness in all three B, V and R bands is reasonable.

Using for example the point source 90$\%$ completeness levels of A06a
(applicable to GCs, which are nearly point sources at the distance of
Coma), we have complete V band data down to V=24--24.25, about half a
magnitude deeper than the bright end of the GC LF.

R band data are deeper than this for nearly all object types.
However, B band data are only complete down B=24.75. So, for objects
with B-V greater than 0.75 (typically elliptical like objects), we can
estimate the GC percentages down V$\sim$24. For B-V greater than 0.5
(typically spiral like objects), we can estimate the GC percentages
down to V$\sim$24.25. For these reasons, not all regions provided a GC
percentage: we limited our analysis to the regions where more than 15
objects were statistically in Coma, detected both in B, V and R and
fainter than V=23.75 (the bright end of the GC LF).

For magnitudes fainter than B=24.75, we have no way to estimate
unambiguously what will be the GC contamination in our LFs using
colors. We know, however, that given the GC physical sizes, such
objects will be unresolved in our data. Objects significantly larger
than the seeing size are therefore likely to be galaxies rather than
GCs (unless several GCs are close to each other in order to mimic
objects larger than the seeing). If we consider the R band major axis
of objects fainter than B=24.75, we find that 53$\%$ of these objects
are two times larger than the seeing in R. This means that a
significant part of the very faint object population in Coma (B fainter than
24.75) is unlikely to be GCs.

Using a bootstrap technique with 500 resamplings, we also
estimated the 1$\sigma$ level uncertainty on the total number of
objects (see Table~\ref{tab:GCs}). If the percentage of GCs is smaller
than three times this uncertainty, we considered the GC population as
not significant. We clearly see that only subfields 8 and 15 have a
potentially significant GC population.

We also independently searched for regions with a significant GC
percentage by summing all the densities given by Mar\'\i n-Franch $\&$
Aparicio (2003). Among our subfields 8, 9, 13, 14, 18 and 19, only
subfield 8 has a significant GC population (at a level larger than
3$\sigma$): these authors found a GC density of 0.0545$\pm$0.018 in
the vicinity of NGC~4874. Subfield 8 mainly includes this galaxy and
is very close to the field surveyed by Bernstein et al. (1995), who
also concluded that there was a significant population of GCs.  

\section{Luminosity function characteristics}

We first note that the general shape of the bright part of the galaxy LF is
similar over the whole field. Besides the dips discussed later, the
faint LF parts (typically R fainter than 19) show spatial variations
discussed in this section.

\subsection{Luminosity function across the Coma field}

In order to sample possible environmental effects on the LF, we cut
the Coma field in twenty 10$\times$10~arcmin$^2$ areas. Results for
the corresponding LFs are shown in Figs.~\ref{fig:fdl1},
\ref{fig:fdl2}, \ref{fig:fdl3} and \ref{fig:fdl4} for the four
available bands (computed with the simple statistical subtraction
method).

\begin{figure*}
\centering
\caption[]{B band LFs for different locations in the Coma
cluster. North is top and East is left. The locations of the three
main galaxies (NGC~4874, NGC~4889 and NGC~4911) are also indicated,
together with the total number of galaxies and with the number of
galaxies brighter than R=23.5 inside each
line of sight. The empty sub-graphs correspond to areas where we did
not have B and V data for technical reasons. The red solid line
separates the north-northeast and south-southwest regions. Continuous
lines correspond to the F02 comparisons and dotted lines to the F10
comparisons. The horizontal lines in each subgraph are only given
as a visual reference to be able to compare more easily results
between the different regions.}
\label{fig:fdl1}
\end{figure*}

\begin{figure*}
\centering
\caption[]{Same as Fig.~\ref{fig:fdl1} for the V band.}
\label{fig:fdl2}
\end{figure*}

\begin{figure*}
\centering
\caption[]{Same as Fig.~\ref{fig:fdl1} for the R band.}
\label{fig:fdl3}
\end{figure*}

\begin{figure*}
\centering
\caption[]{Same as Fig.~\ref{fig:fdl1} for the I band.}
\label{fig:fdl4}
\end{figure*}

For all photometric bands and individual exposures, there
is clearly a dichotomy between the north-northeast regions, with a
steeply rising LF, and the south-southwest regions with flatter
LFs. This could be due to different galaxy types which behave
differently. In order to address this observation, we used the
color-dependent statistical subtraction method in two different
ways.

In a first approach, we compute LFs for blue and red galaxies selected
using the CMR Red Sequence (hereafter RS) defined in A06a. This RS has
a given width and we considered as blue the galaxies statistically in
Coma and under the RS at the significance level of 1$\sigma$. 
Other galaxies statistically in
Coma were considered as red. This separation corresponds at first
order to a classification into early and late type galaxies
(e.g. Fukugita et al. 1995). The results are given in
Fig.~\ref{fig:fdlbrcol} and we clearly see that the blue object fraction 
increases toward faint magnitudes. This is in good agreement with the 
results by Odell et al. (2002).
Fig.~\ref{fig:fdlbrcol} is, however, not directly comparable to the R band LFs
given in Fig.~\ref{fig:fdl3}. This is because it uses the color-dependent 
statistical subtraction method with B and R while Fig.~\ref{fig:fdl3} uses the 
simple statistical subtraction method. The B band data are not
as deep as the R band data: a galaxy has to be detected both in B and
R to be included in the color LF. Relatively faint and very
low star-forming rate objects will be excluded because they are
fainter in the B band than in the R band.

\begin{figure*}
\centering
\caption[]{Same as Fig.~\ref{fig:fdl1} for red galaxies (black/red
lines) and blue galaxies (grey/green lines). Error bars are not shown
for clarity.}
\label{fig:fdlbrcol}
\end{figure*}

\begin{figure*}
\centering
\caption[]{Same as Fig.~\ref{fig:fdl1} for early type galaxies
(black/red lines), early spiral galaxies (light grey/green lines) and
late spiral galaxies (heavy black/blue lines). Error bars are not
shown for clarity.}
\label{fig:fdlbricol}
\end{figure*}

As a second approach we use synthetic colors, which provide a robust
classification of galaxies into spectrophotometric types.  The lack of
deep U band data prevents any reliable computation of photometric
redshifts in Coma (e.g. Savine 2000) but a color classification of
galaxies is still possible using statistical field subtractions in the
color-space. Galaxy templates from Coleman et al. (1980) were used to
derive the expected colors of galaxies at the redshift of Coma and
discriminate between different (extreme) galaxy types in the
color-space, from E to Im galaxies.  Different simulations were
performed to estimate the contamination level expected between
adjacent types when including realistic photometric uncertainties. As
a result, we define three galaxy populations, separated in the B-I/B-R
space by two lines:

(B-I) = $-$1.2(B-R) + 1.45

\noindent
and

(B-I) = $-$1.2(B-R) + 2.60

This classification roughly corresponds to a separation between early
type, early spiral and late spiral galaxies. With these limits, only
15$\%$ of the Coleman et al. (1980) type 2 (Sbc) are misclassified as
type 1 (E/S0) at the typical S/N levels of the faintest
galaxies in the sample. Similarly, only 26$\%$ of type 3/4 galaxies
(Scd/Irr) are classified as type 2.

The first and second approaches are complementary. While the CMR only
depends on the early type galaxy observed colors, the synthetic color
modelling depends on the templates used. The second method is
more predictive (subdivision into three galaxy classes instead of two)
but also more model-dependent. The fact that these two methods provide
similar results gives us confidence in our results.

Whatever the galaxy colors or types, the LFs in the south-southwest
regions remain flat (see Fig.~\ref{fig:fdlbricol} for the three
populations defined in the color-space), except perhaps at the faint
end for the early and late spiral galaxies. The situation is more
complex in the north-northeast parts of the cluster. Red galaxies have
a similar behaviour than early type galaxies. Early spirals clearly
dominate late spirals (as expected in a massive cluster) except
perhaps in the western parts such as subfield 5 where the numbers of
early and late spirals are comparable. The steep LF rise is due to
red/early type galaxies for R$<$22.5 and to blue/spiral galaxies
after.

\subsection{Faint Coma cluster luminosity function around the main 
cluster galaxies}

We now describe the LF shapes, which are quite different, around the
main cluster galaxies and groups in the Coma field, i.e.: NGC~4874,
NGC~4889 and NGC~4911 (see Figs.~\ref{fig:fdl1}, \ref{fig:fdl2},
\ref{fig:fdl3} and \ref{fig:fdl4}). NGC~4911 is a well known galaxy
attached to an infalling group (e.g. Neumann et al. 2003) while
NGC~4874 and NGC~4889 are the two central galaxies of the cluster.
Note that we are not sampling in our data the NGC~4839 region, another
infalling group.

In the NGC~4889 field, LFs are steeply rising in all photometric bands
up to B$\sim$23.5, V$\sim$23, R$\sim$23 and I$\sim$23, then decrease
above these limits in the B and V bands, while continuously rising or
at least staying flat in R and I .

In the NGC~4874 field, the LF is moderatly rising in B and V with a
steeper rising in R and I, and a maximum reached around R$\sim$23 and
I$\sim$22.5.

In the field of NGC~4911, the LF is moderately rising in B and V up to
B$\sim$23.75 and V$\sim$23.5. In R and I, the LF is oscillating but
still rising up to R$\sim$23 and over the whole magnitude range in I.

We will discuss in the following the link between these shapes and the
general cluster building history.

\begin{table}
\caption{Schechter function parameters for the north-northeast (N) and
south-southwest (S) fields.}
\begin{tabular}{lrrrr}
\hline
 & B~~~ & V~~~ & R~~~ & I~~~ \\
\hline
N $\alpha$ & -1.48$\pm$0.03 & -1.72$\pm$0.01 & -1.74$\pm$0.02 &
-1.60$\pm$0.01 \\
N M$^*$ & 14.89$\pm$0.46 & 12.90$\pm$0.24 & 14.68$\pm$0.91 & 12.31$\pm$0.23 \\
S $\alpha$& -1.32$\pm$0.03 & -1.27$\pm$0.03 & -1.28$\pm$0.05 &
-1.27$\pm$0.02 \\
S M$^*$& 15.86$\pm$0.44 & 14.82$\pm$0.99 & 16.05$\pm$0.36 & 14.29$\pm$0.25 \\
\hline
\end{tabular}
\label{tab:parameters}
\end{table}

\subsection{Modelling the Coma cluster luminosity function}

We will now give a Schechter modelling of our LFs. This is not a
crucial step in our study as we do not use this model, but it can be
useful for readers in order to compare our results with other studies
in the literature.  We used a single Schechter model of our LFs fitted
to the data using an IDL code already discussed in Durret et
al. (2002). However, given the clearly non-Schechter LF shapes (with
many bumps and dips) the results are only indicative. The less
constrained parameter is M$^*$, while the faint magnitude 
slope of the Schechter function is better constrained. This
induces degeneracies between the two parameters but we did not try to
perform a finer analysis given the limited goals of this modelling.

\begin{figure}
\centering
\mbox{\psfig{figure=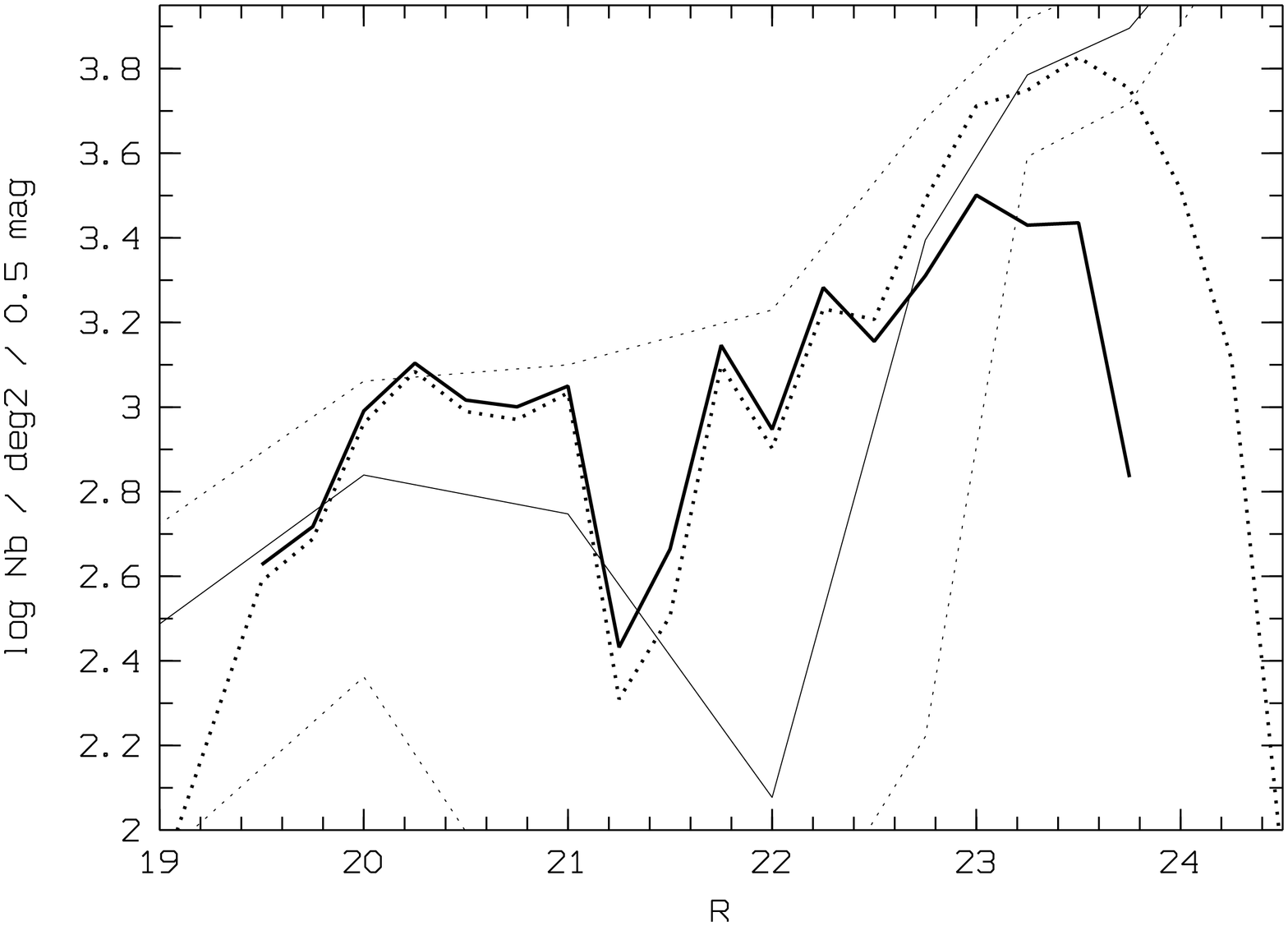,width=8cm,angle=270}}
\caption[]{Comparison with Bernstein et al. (1995: R band). Our own LF
estimates are the thick lines, the solid line is the F02
comparison field, and the dotted line for the F10 comparison
field. The thin solid line is the literature estimate along with its
1$\sigma$ errors (thin dotted lines).}
\label{fig:compar11}
\end{figure}

\begin{figure}
\centering
\mbox{\psfig{figure=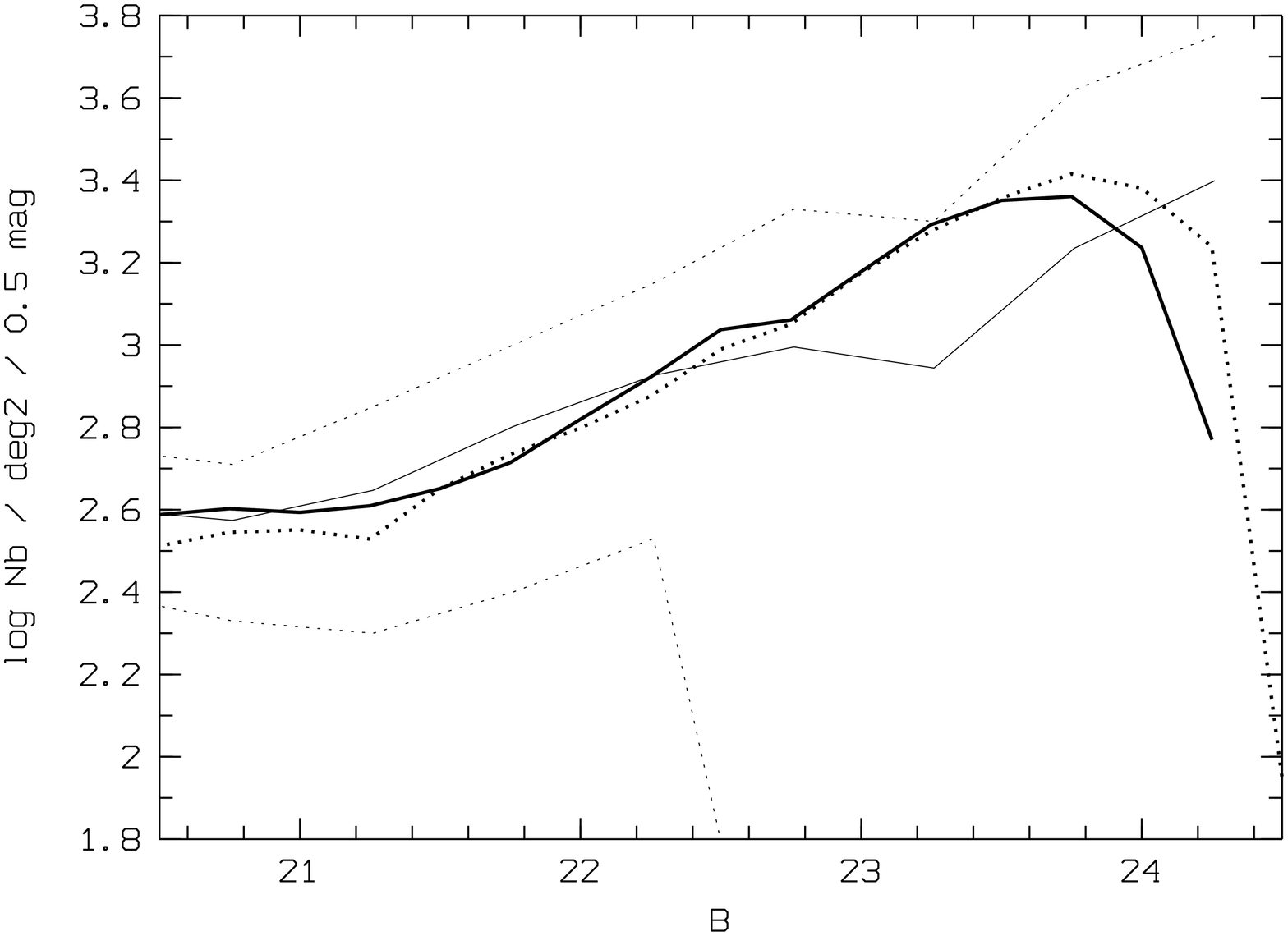,width=8cm,angle=270}}
\mbox{\psfig{figure=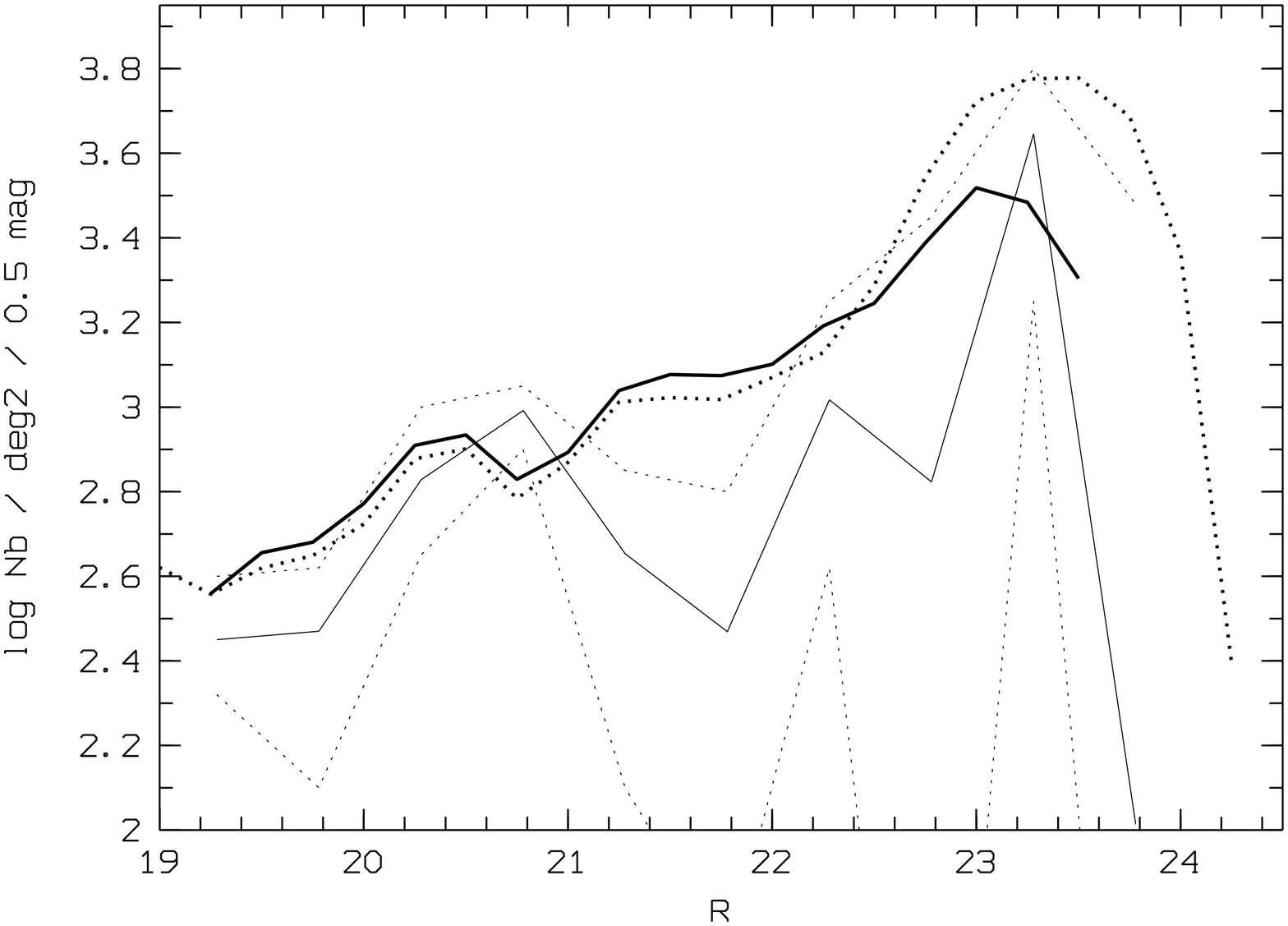,width=8cm,angle=270}}
\caption[]{Same as Fig.~\ref{fig:compar11}: comparison with Trentham
(1998); from top to bottom: B and R bands.}
\label{fig:compar12}
\end{figure}

As shown in the last subsection, we can divide the Coma cluster in two
parts. A north-northeast (fields 20, 19, 18, 17, 15, 14, 13, 10, 9, 8,
5 and 4) and a south-southwest part (fields 16, 12, 11, 7, 6, 3, 2 and
1). For simplicity we will call these fields the north and the south
fields respectively. LFs for these two regions as a whole are shown in
Figs.~\ref{fig:fdlnord} and \ref{fig:fdlsud} and the modelling results
are given in Table~\ref{tab:parameters}. North is modeled by a
Schechter function with a steep slope due to red/early type galaxies
at bright magnitudes, and to blue/late type objects at very faint
magnitudes, while on the contrary, the south LF is nearly flat for all
objects. The values of M$^*$ are very poorly constrained and
their values in B, V and R for the north and south regions are
consistent within error bars. However, we can note that the south
values are always fainter than in the north (the difference is
significant at the 3$\sigma$ level for the I band). We will not
discuss further this point as uncertainties are very large and LF
shapes are clearly not Schechter-like.

\subsection{Comparison with literature luminosity functions}

We present in this section a short comparison of our Coma LFs
derived from the simple statistical subtraction method (without use of
color information) with those derived by Bernstein et al. (1995), Lobo
et al. (1997), Trentham (1998), Andreon $\&$ Cuillandre (2002),
Beijersbergen et al. (2002) and Iglesias-P\'aramo et al. (2003). These
authors also use statistical background galaxy subtraction methods but
most of the time in smaller fields of view or with limited depth or
less detailed spatial LF analysis.

LFs by Bernstein et al. (1995) in the R band, Trentham (1998) in the B
and R bands and Beijersbergen et al. (2002) are given for areas
included in our data, but with an overlap of less than 2 magnitudes in the
case of Beijersbergen et al. (2002). We therefore computed our LFs in
the corresponding bands only for the Bernstein et al. (1995) and
Trentham (1998) fields. Results are shown in 
Figs.~\ref{fig:compar11} and \ref{fig:compar12}.

Given the error bars, the agreement is globaly good in all bands
except that the dip (at R$\sim$22) visible in the Bernstein et
al. (1995) LF is less prominent in our results. This is probably due
to the small comparison area used by these authors.

We have only partial overlaps with Lobo et al. (1997), Andreon $\&$
Cuillandre (2002) and Iglesias-P\'aramo et al. (2003) and only checked that
we also have a reasonable agreement with their results.

\section{Red-Sequences in the Coma cluster}

Using color dependent statistical subtractions naturally leads to
investigate the possible existence of a RS at faint magnitudes in the
Coma CMR (B-R vs. R). We will describe here the origin of this RS and the
consequences of its characteristics.

\subsection{Why a Red-Sequence in the Coma cluster?}

The existence of a correlation between early type galaxies and their
colors has been known for almost 50 years (e.g. Baum 1959): elliptical
and S0 galaxies seem distributed along a RS. This RS proved to be even
more obvious in clusters of galaxies, where galaxy evolution is
different from that in the field because of environmental
effects. Sandage (1972) has shown for example the existence of a very
clear RS for the elliptical galaxies of the Virgo and Coma clusters.
The use of the RS of early type galaxies in the CMR rapidly became a
common tool to study galaxy clusters (e.g. Mazure et al. 1988, Yee et
al. 1999). There is also a visible sequence for late type and dwarf
galaxies (Terlevich et al. 2001) but this sequence is more dispersed
and shows different colors compared to the elliptical galaxy RS.

The RS in the CMR is mainly due to metallicity effects in galactic
systems of various masses (Kodama $\&$ Arimoto 1997). At the Coma
redshift, the B-R color is a good tracer of the ratio between the
galactic star-forming rate and the galactic mass and therefore the RS
traces the star formation history of these galaxies and is directly
related to the metal abundances. The more massive a galaxy, the more easily
it will retain metals against dissipative processes such as
supernova winds. The more metals present in a galaxy, the reddest the
galaxy will be. This is a simple way to account for the negative slope
of the RS in clusters of galaxies.

These characteristics are well established for relatively bright
galaxies but little is known for low mass galaxies beyond the
spectroscopic limit. Authors as Terlevich et al. (2001) already
extensively discussed the Coma CMR characteristics, but these previous
studies typically end where our sample starts, and the magnitude range
beginning at M$_R$$\sim -14$ basically remains an unexplored domain
regarding the CMR. Even deep spectroscopic surveys (e.g. Adami et
al. 2000 or A06a) were not able to significantly sample the CMR in
Coma at magnitudes fainter than R$\sim$21. However this range is of
major interest because it contains low mass galaxies with various
histories and for example we showed in Adami et al.  (2006b) that
faint low surface brightness galaxies in the Coma cluster are not all
located on the RS.

We will now investigate the CMR in the Coma cluster down to
M$_R$$\sim -11.5$.

\subsection{Removal of field galaxies}

As described earlier, in order to remove field galaxies, we computed
density maps of the Coma line of sight and of the empty field in the
B-R / R space. This was achieved via kernel techniques (e.g. Biviano
et al. 1996) providing a good compromise between space sampling and
resolution of the maps.

We then renormalized the two maps to the same area, before subtracting
the empty field map to the Coma line of sight map. The resulting map
is therefore the B-R / R distribution for objects statistically in
Coma (see Fig.~\ref{fig:mosaic}).

\subsection{Significance of the B-R/R maps}

In order to interpret the maps given in Fig.~\ref{fig:mosaic}, we must
assess their significance level. The galaxy density D in these B-R/R maps is
computed as the difference between the galaxy density along the Coma line of
sight (D$_{C~los}$) and along the empty field (D$_{E~f}$):

D(B-R,R) = D$_{C~los}$(B-R,R) - D$_{E~f}$(B-R,R)

We used 500 bootstrap resamplings to compute uncertainties on D$_{C~los}$ and
D$_{E~f}$. The total uncertainty on D is the quadratical sum: 

$\sigma$$^2$$_D$(B-R,R) = $\sigma$$^2$$_{C~los}$(B-R,R) + $\sigma$$^2$$_{E~f}$(B-R,R)

We finally compute the significance level maps (Fig.~\ref{fig:mosaicsig})
dividing D(B-R,R) by $\sigma$$_D$(B-R,R). We see that most of the
  populated regions in Fig.~\ref{fig:mosaic} are significant at more than the
  3-$\sigma$ level according to Fig.~\ref{fig:mosaicsig}. Some regions where
  there is more objects in the empty fields compared to the Coma line of
  sight or where the significance level is very low also appear in 
  Fig.~\ref{fig:mosaicsig}. They mainly
  affect some of the Coma south fields in faint and blue areas (and a few bright
  regions obviously not populated as shown in Fig.~\ref{fig:mosaic}).

\subsection{Global distribution in the B-R/R space}

In Fig.~\ref{fig:mosaic} we show the areas where we detect galaxies
statistically belonging to the Coma cluster, as a function of location
in the cluster. For comparison, the RS computed for galaxies brighter
than R$\sim$18.5 in A06a is overplotted on these graphs. Above
R$\sim$19.5 where there is only incomplete or no spectroscopic
information we can clearly see positive and significant 
galaxy concentrations, most of the time following the bright galaxy 
RS, whatever the location in the cluster. This is in good agreement 
with the results of Odell et al. (2002) showing a linear RS from $-13$ to 
$-22$ in a narrow band filter.

In addition to the RS, extensions towards the red parts of the B-R/R
plots show an excess of very red objects (significant and distinct 
from the ones we will discuss in Section 8.4) along the Coma cluster line of
sight compared to the field. These objects are so red (B-R$\sim$2.5)
that they are difficult to explain as part of the Coma cluster itself,
because even very red early type galaxies are most of the time not so
red (e.g. Fukugita et al. 1995). Only atypical dust contents can
provide such red colors at the Coma cluster redshift. Moreover, this
contribution is not negligible and objects with B-R greater than 2
account for 19$\%$ of all the objects potentially belonging to the
Coma cluster. This percentage does not vary strongly across the field,
and range typically from 14 to 23$\%$. An explanation could be the
presence of more distant concentrations, distinct from Coma, between
z$\sim$0.16 and 0.18 (see A06a) and spread over the whole Coma field.

In order to check this hypothesis, we translated the Coma cluster RS
to a 0.17 redshift using the elliptical galaxy synthetic models of
Coleman et al. (1980). The new RSs obtained are shown in
Fig.~\ref{fig:mosaic} as the green/grey lines and reasonably
correspond to the locations where an excess of red objects is
detected.

Finally, by simply counting the number of spectroscopic redshifts
available between z$\sim$0.16 and 0.18 in the Coma field, we find a
number of redshifts equal to $\sim$15$\%$ of the total number of
redshifts inside the Coma cluster itself, in good agreement with the
percentages of very red objects previously found. These galaxies
spectroscopically located at z$\sim$0.17 extend all across the field
and are probably related to a distant filament on the line of sight
that we will not study here. This background structure is probably the
same as the one at $\sim$49000 km/s quoted by Guti\'errez et
al. (2004), with a velocity dispersion of 1000 km/s).

In conclusion, the population of objects fainter than R=19.5 seen in
excess in the Coma field compared to the empty fields is well
explained by the contributions of objects inside the Coma cluster
globally following the bright Coma cluster galaxy RS and of a 
population of very red objects at z$\sim$0.17 that we will not consider in the
following.

\begin{figure*}
\centering
\caption[]{CMR for different locations in the Coma cluster. North is
top and East left. Locations of the three main galaxies (NGC~4874,
NGC~4889 and NGC~4911) are shown. The red/black inclined line is the
RS computed in A06a using bright galaxies with spectroscopic redshifts
inside the Coma cluster. The green/grey inclined line is the same RS
shifted to z=0.17 using synthetic elliptical galaxy templates. The
black/blue rectangles at faint magnitudes are the locations where
intergalactic globular clusters can pollute the galaxy sample. The red
broken line delineates the north and south fields. We also give for
each subfield the mean CMR slope along with the subfield number.}
\label{fig:mosaic}
\end{figure*}

\begin{figure*}
\centering
\caption[]{Same as Fig.~\ref{fig:mosaic} with different colors
according to significance levels of densities given in Fig.~\ref{fig:mosaic}. 
Black shaded regions: between 1 and
3$\sigma$ levels, red/light-black shaded areas, between 3 and
4$\sigma$ levels, green/grey shaded regions: above 4$\sigma$
levels. The white areas embedded in colored areas are significant
at less than 1$\sigma$ (or even regions where empty fields dominate the Coma
line of sight population). The blue
inclined thick lines show the red sequence of bright galaxies computed
in A06a.}
\label{fig:mosaicsig}
\end{figure*}

\subsection{Location of Coma cluster objects in the B-R/R relation
relatively to the RS of bright galaxies}

\begin{figure}
\centering
\mbox{\psfig{figure=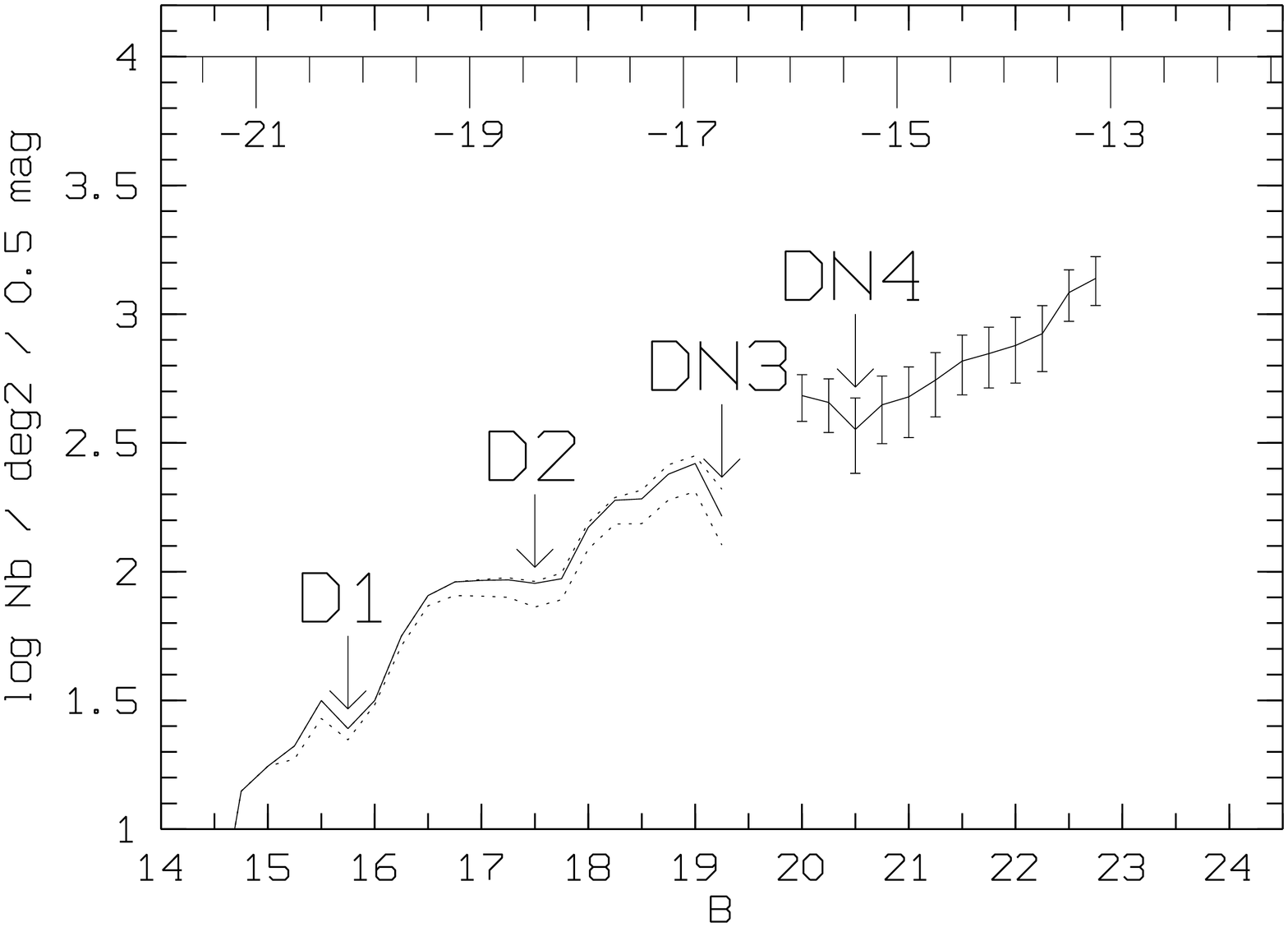,width=8cm,angle=270}}
\mbox{\psfig{figure=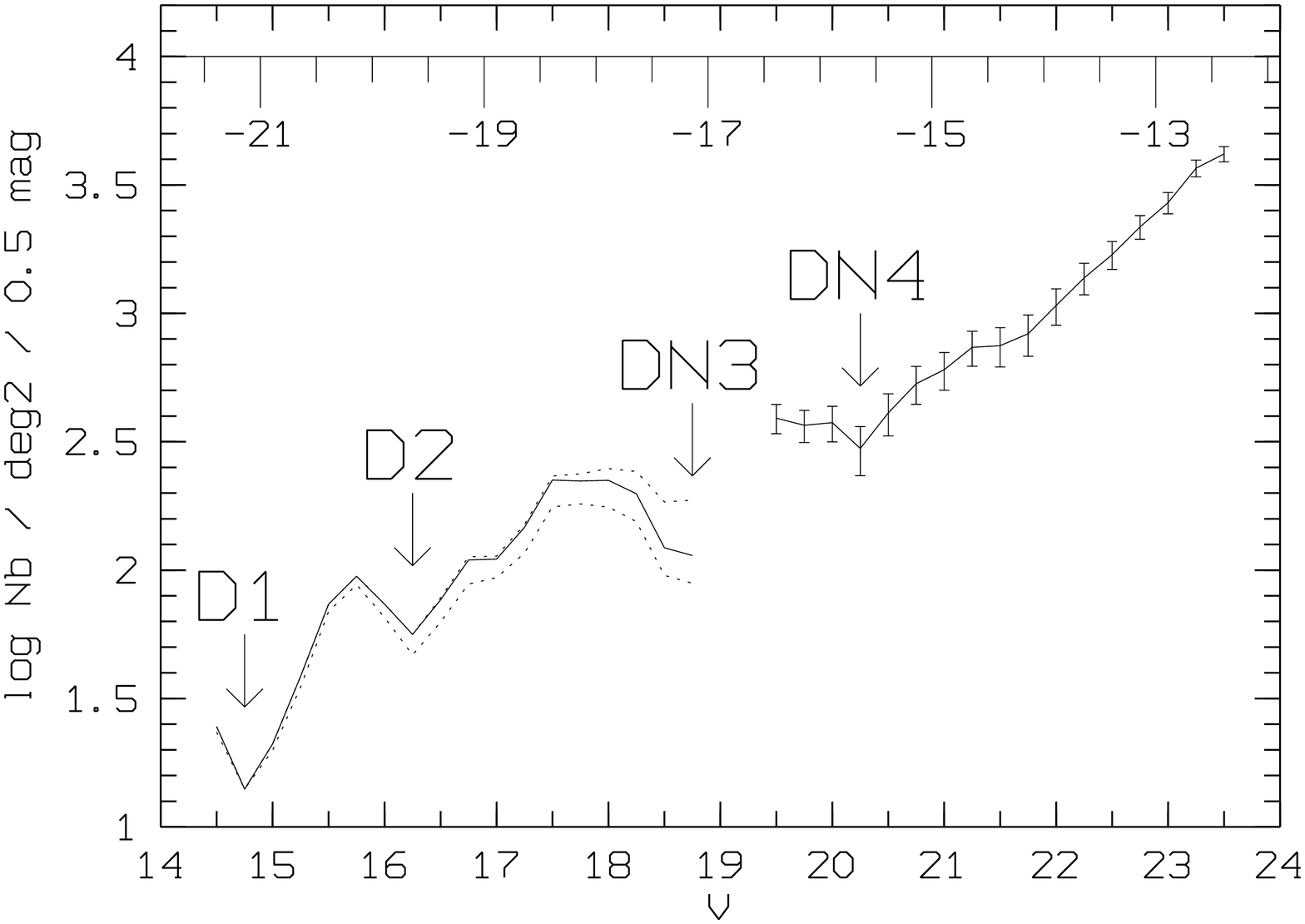,width=8cm,angle=270}}
\mbox{\psfig{figure=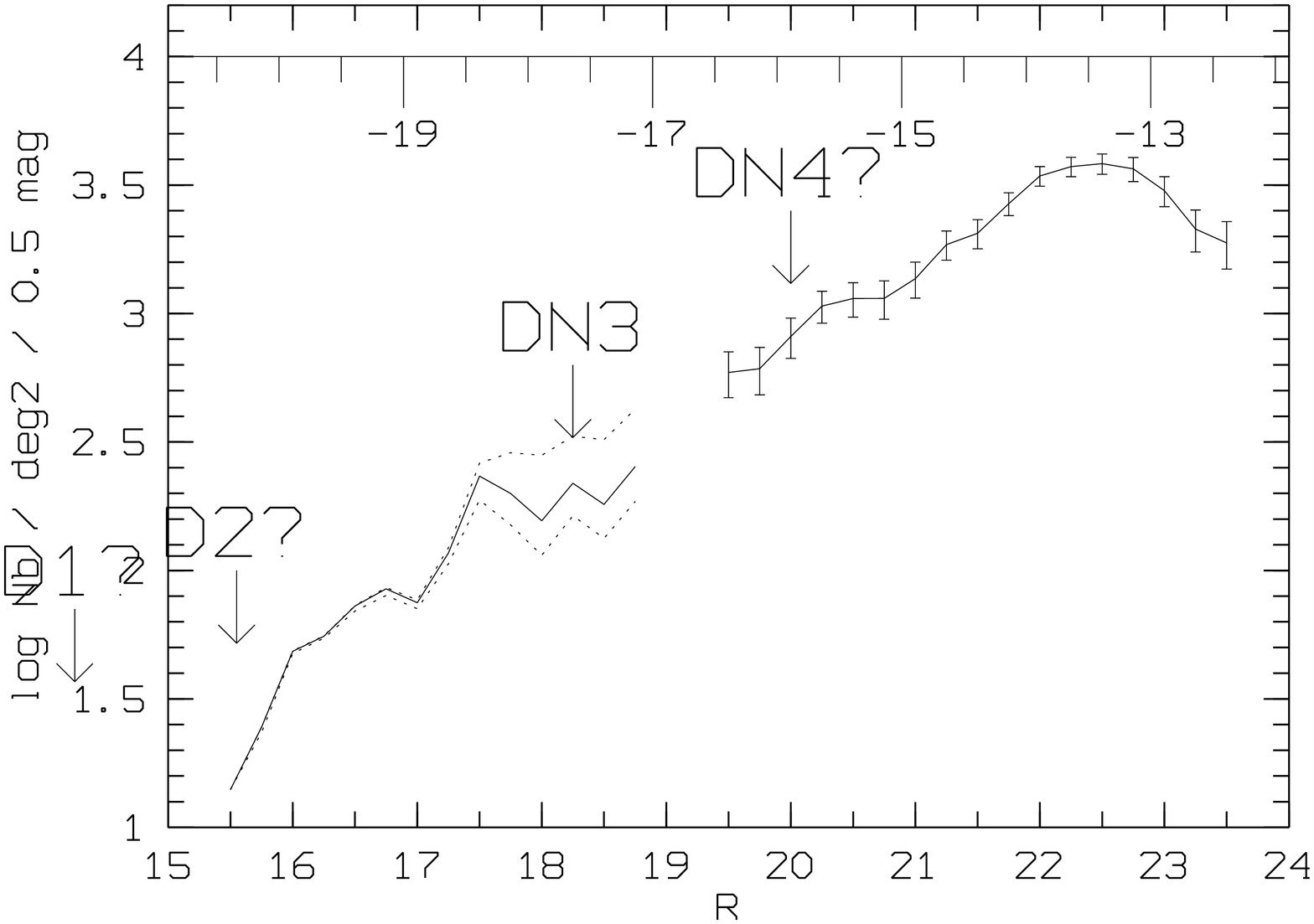,width=8cm,angle=270}}
\mbox{\psfig{figure=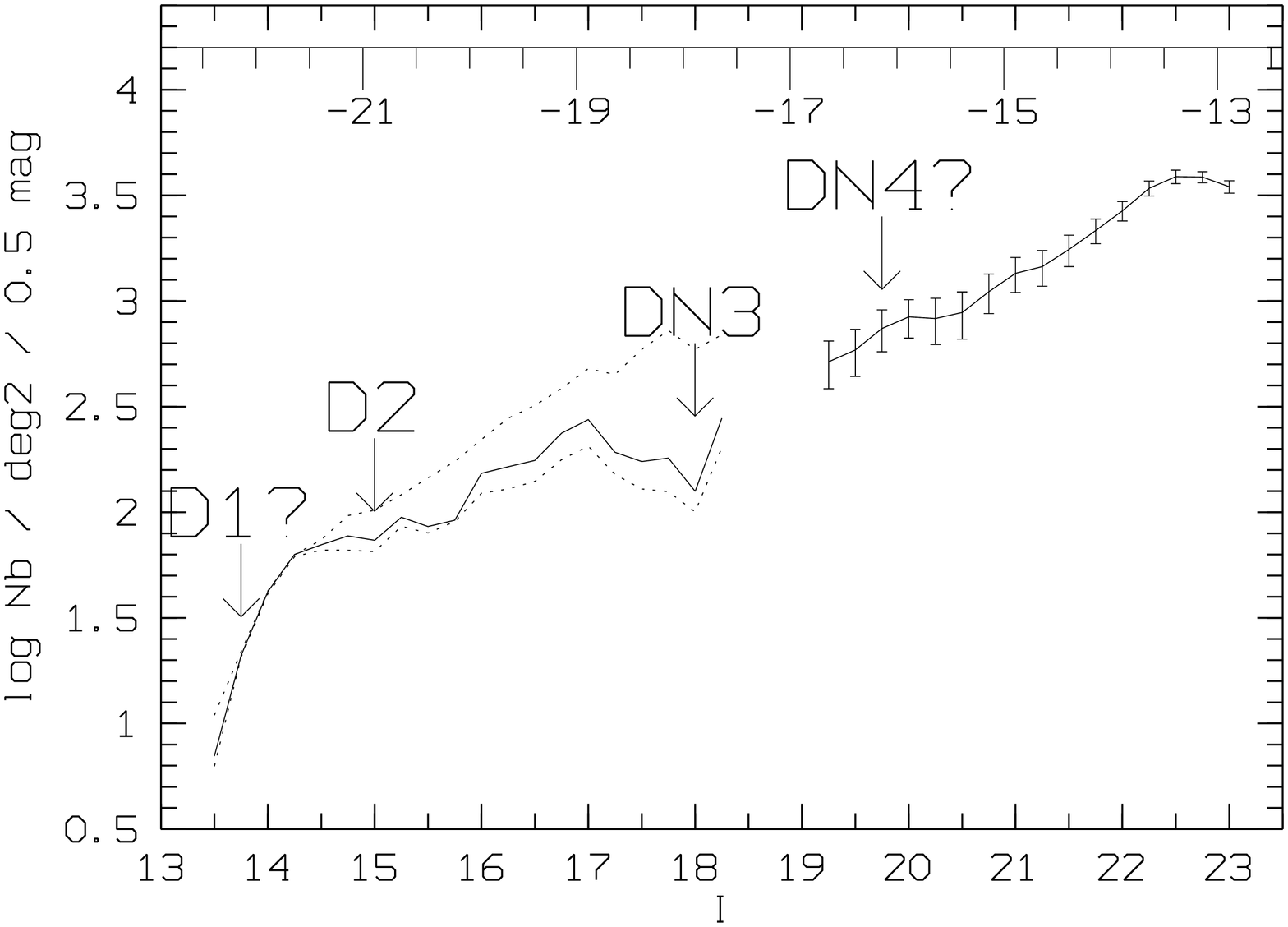,width=8cm,angle=270}}
\caption[]{BVRI LFs for the north-northeast area. We give both 
apparent and absolute magnitudes. The continuous line on the left half
is the LF built from spectroscopy with its 1$\sigma$ uncertainty
envelope (dotted lines). The continuous line to the right is the LF
built from statistical subtraction without any use of color
information. Error bars come from the methods explained in the text. The
position of the dips in all colors is also given.}
\label{fig:fdlnord}
\end{figure}

\begin{figure}
\centering
\mbox{\psfig{figure=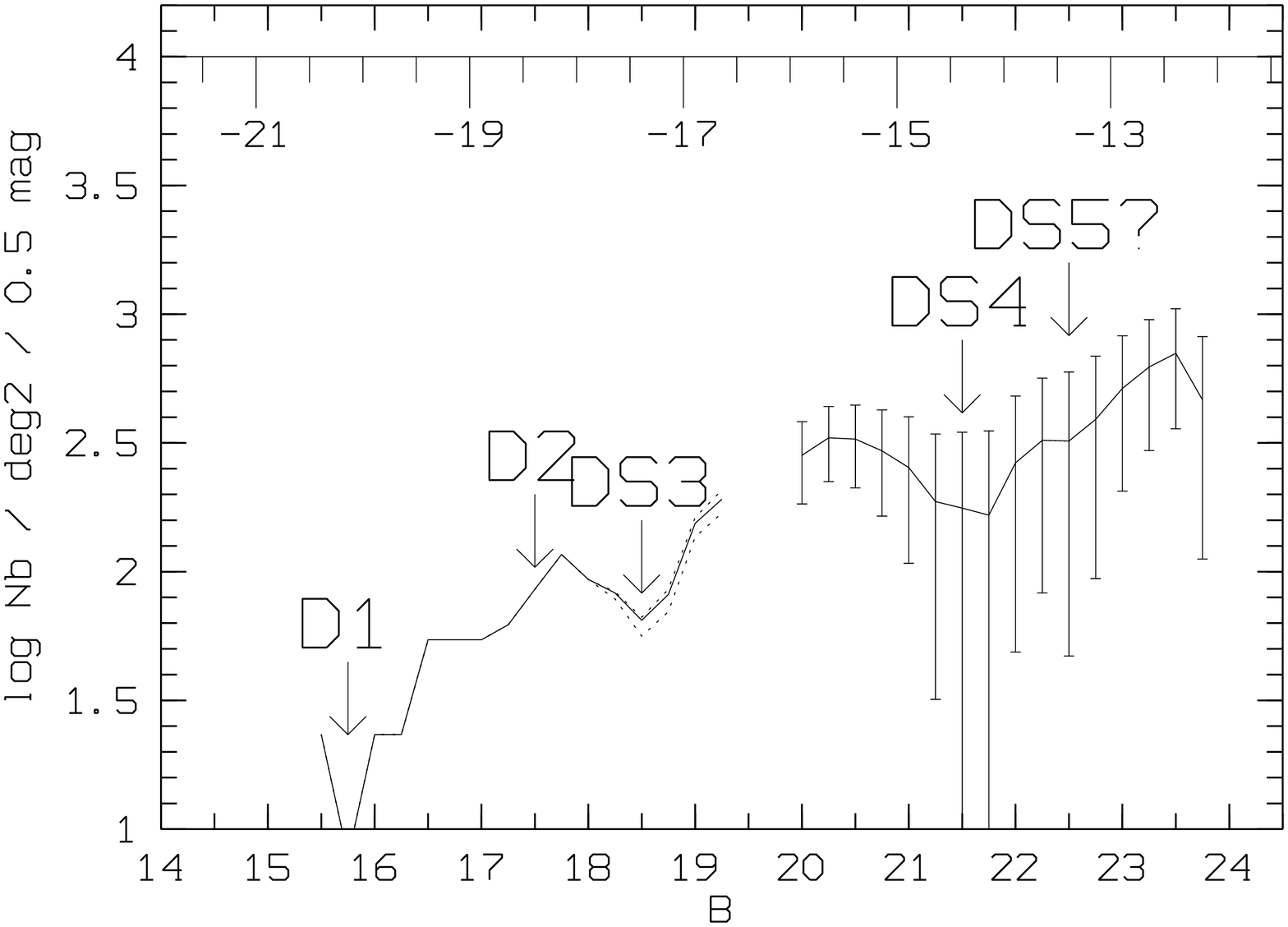,width=8cm,angle=270}}
\mbox{\psfig{figure=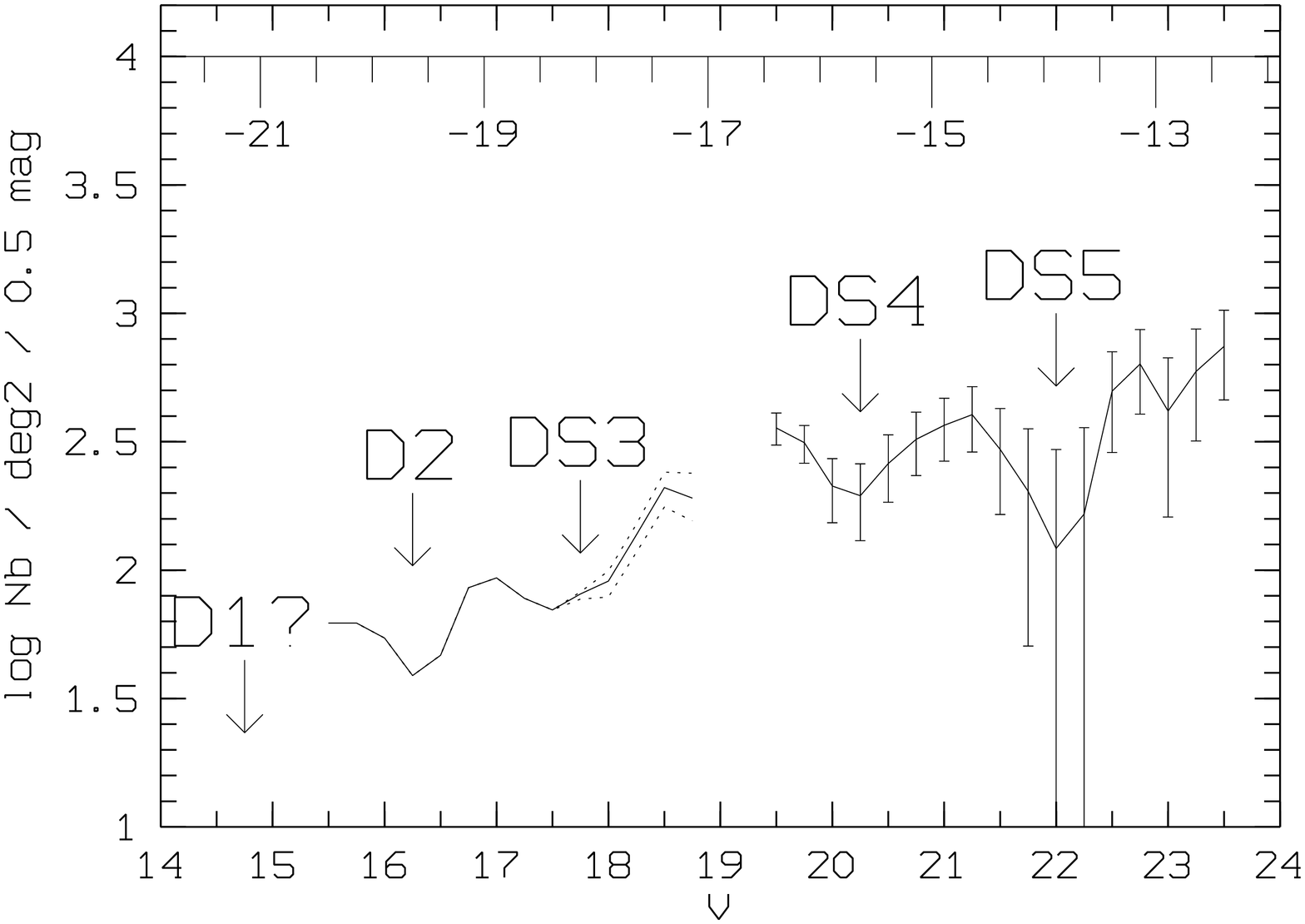,width=8cm,angle=270}}
\mbox{\psfig{figure=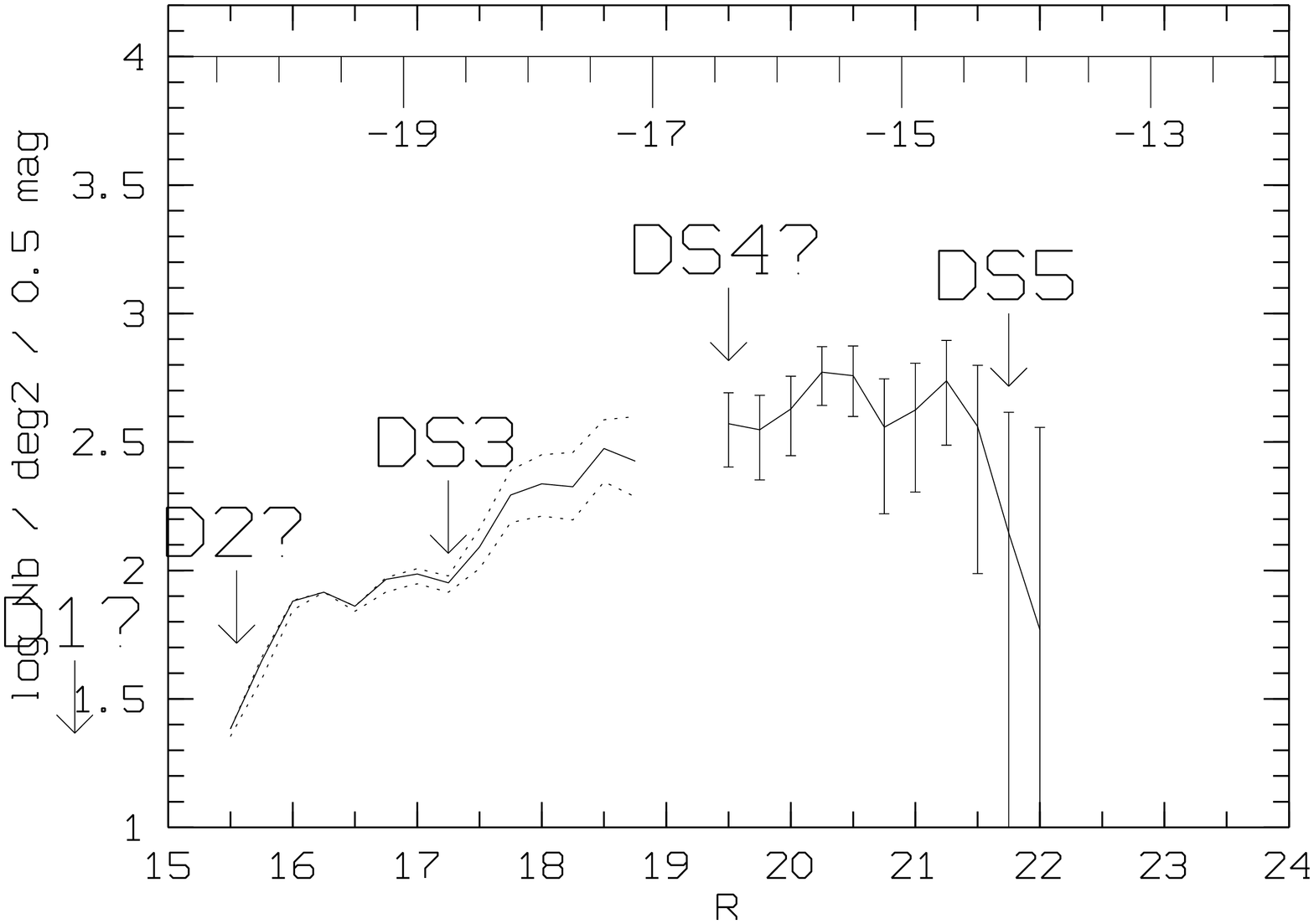,width=8cm,angle=270}}
\mbox{\psfig{figure=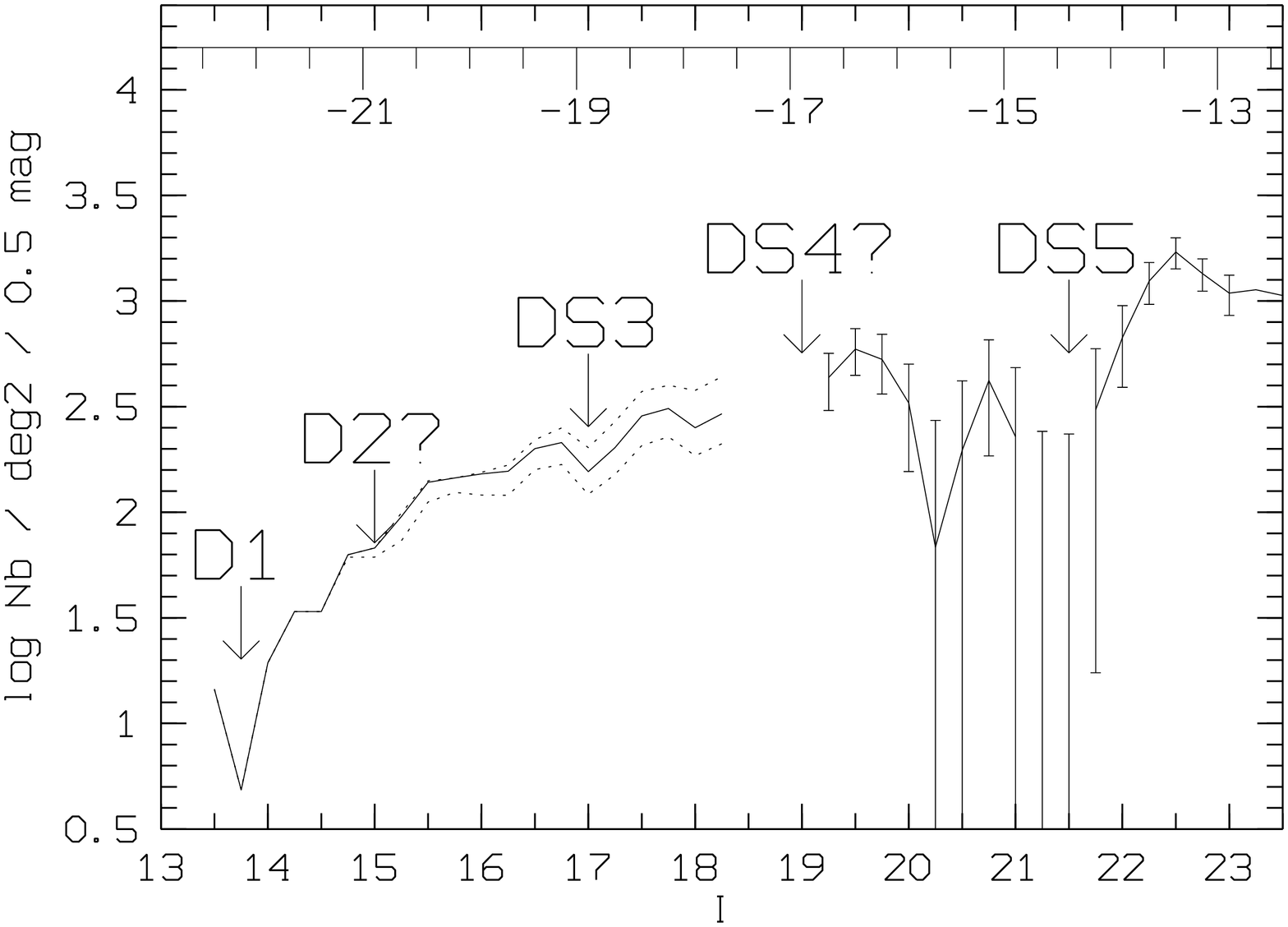,width=8cm,angle=270}}
\caption[]{Same as Fig.~\ref{fig:fdlnord} but for the south-southwest region.}
\label{fig:fdlsud}
\end{figure}

Looking at Coma cluster objects in more detail, we can also see in
Fig.~\ref{fig:mosaic} that there are two populations close to but outside of 
the RS of bright galaxies: one being redder (distinct from the z$\sim$0.17
very red population previously quoted) and the other bluer, with
distributions depending on the location in the field. We will refer to these
two populations in the following as the red and blue galaxies.

The south zone exhibits significant (see
Fig.~\ref{fig:mosaicsig}) red object concentrations at R$\sim$22-23
while the north zone seems to show fewer red objects. We also
have a significant blue population in the north. This blue
concentration is also visible in the south regions but is not
always significant. 

We also note some regions in the south where empty fields galaxy populations
dominate the Coma line of sight. This can be due to local overestimate of the
background galaxy population but is perhaps also due to merging/disruption
events inside the Coma cluster, removing galaxies from the counts.

There is a variation of colors as a function of location in the
  cluster. This is visible on the mean slope of the CMR as a function of
magnitude (see Fig.~\ref{fig:mosaic}). The south shows a mean negative
slope of $-0.044 \pm 0.014$ (characterizing the whole Coma cluster population, 
not only the early type galaxies included in the bright CMR RS)
while the North exhibits a slope of $-0.084 \pm 0.031$.

\subsection{Contribution of intergalactic globular clusters?}

As explained earlier, our data are by definition free from GCs still
linked to giant galaxies.  However, we may still have detected
intergalactic GCs or GCs attached to faint galaxies even if this
contribution is probably minor (Mar\'\i n-Franch $\&$ Aparicio 2003). 

Nevertheless, the brightest magnitudes of GCs are not bright enough to
completely account for the red objects between R=22 and 23 in the
south field. GCs could contribute to the excesses of blue objects in
the north fields at R fainter than 23 and B-R$\sim$0.7, but there should 
be only a few GCs existing with such blue colors (e.g.
Rhode $\&$ Zepf 2004) at these magnitudes and this contribution should
be minor.

In conclusion, GCs do not allow to account for the red and blue object 
excesses statistically detected in the Coma cluster. A significant
part of these blue objects must therefore be galaxies.

\section{Inferences from the Coma LF on the cluster building up 
scenario}

\subsection{Previous Coma cluster literature luminosity functions and
 environmental effects}

In their pioneering work, Bernstein et al. (1995) analyzed the Coma LF
in a small but very deep field close to NGC~4874 and found a faint end
Schechter slope of $\alpha=-1.42$.  However, the field was too small
to show evidence for large scale environmental effects.

Lobo et al. (1997) were the first to derive relatively deep V band
galaxy LFs in various regions of the Coma cluster.  They showed that
the faint end slope of the LF was flatter in the central regions
($\alpha=-1.51$ to $-1.58$) and steeper in the outskirts (as steep as
$-1.88$), with error bars of $\pm 0.10$. They proposed a dynamical
interpretation based on the recent history of the two subclusters
surrounding the giant galaxies NGC~4874 and NGC~4889: when these two
groups have fallen into the central region of Coma, the large number
of dwarf galaxies initially present may have been rapidly accreted by
the largest ones, therefore flattening the faint end slope of the LF.

Trentham (1998) confirmed that the faint end slope of the overall
LF could be as steep as $-1.7$; however, he did not find any
difference in the faint end slope of the LF between the inner
(r$<200$~kpc) and outer (r$>200$~kpc) regions. Neither did
Iglesias-P\'aramo et al. (2003), who only report a steepening towards
the outskirts significant at the 1$\sigma$ level.

Beijersbergen et al. (2002) found that the LF slope was steeper
towards the blue ($\alpha=-1.54,\ -1.32$ and $-1.22$ in U, B and r
respectively, for the full sample). They also found a stronger
steepening in the U band toward the cluster edges.

Simultaneously, Andreon \& Cuillandre (2002) observed Coma in smaller
but somewhat deeper fields and also found that the slope and shape of
the LF was dependent on colour.

However, all these studies were not well adapted in terms of combined
depth, area and spectral coverage to investigate properly
environmental effects in the Coma cluster. They were able to sample
different cluster regions (but assuming most of the time an
axisymmetric geometry which is not valid here), or different galaxy
magnitude classes, or LFs in different colors, but never all
together. Our data allows to reach such goals simultaneously.

For example, the Coma cluster building process proposed in Adami et
al. (2005b) allows to interpret our results in terms of infalls onto
the cluster coming from well defined directions: northeast-east,
northwest-west and south-west.

We will now show that the GC distribution and luminosity function
shapes are in good agreement with this framework.

\subsection{Spatial distribution of GCs}

The high fraction of GCs found close to NGC~4874 (see also
Mar\'\i n-Franch $\&$ Aparicio 2003) can be explained by assuming that
it is the oldest cluster dominant galaxy, located at the bottom of the
original cluster potential well. When the GC parent galaxies passed
through the inner cluster core (where the tidal interactions are the
strongest), part of their GCs may have been expelled, leading to a
high concentration of GCs in that zone.

The other region showing atypical GC percentages is in the north part
of the cluster (subfield 15), where about half of the objects between
V=23.75 and V=24.25 have colors consistent with GCs. This region does
not contain major galaxies and a possible explanation (assuming that the
GC percentage is correct) is that a passing through galaxy has lost
part of its GC population.

\subsection{Infall directions and luminosity functions}

The literature on Coma cluster LFs sometimes seems to lead to
contradictory conclusions, probably because these studies usually assume
an axisymmetric geometry which is not correct. As we show in the
present paper, the variations are more complex and follow an overall
nearly north/south symmetry and some peculiar processes related to the
bright galaxies.

The south regions showing quite flat LFs are also those located
on the main field infall directions, symbolized for example by the
infalling NGC~4839 group (not sampled by our data): see Lobo et
al. (1997).  The north regions are not subject to major known infalls
and show steeply rising LFs. The LF shapes could then be related to
the presence or not of directions of infall: the north regions being
only poorly fed by external infalls, we search the explanation in the
south infalls.

First, galaxies infalling from the south can enhance the north
population after having crossed the entire cluster. We would need
about 1500 galaxies coming from the south to be located in the north
region. This number is the difference between the north and south LFs
between R=18 and 22 (the faintest magnitude in the south region where
the LF can still be computed with our data). This number is plausible
given the infall rate given in the following (see Adami et al. 2005b),
but we clearly need simulations to assess this process.

Second, theinfall of galaxies from the south can also diminish the south 
galaxy population by several processes:

- Merging between faint galaxies to form brighter objects cannot work, because 
in this scenario we should have a significant overabundance of bright 
galaxies in the south compared to the north and this is not the case. 

- Merging of faint galaxies with brighter objects could be another
explanation, since this process is independent of galaxy type and
therefore of galaxy color.  Fig.~\ref{fig:fdlbrcol} shows that all the
regions in the south exhibit nearly flat LFs for both blue and red
objects, so the process removing faint galaxies from the counts has to
be exterior to the galaxies, such as for example a fusion scenario.

Adami et al. (2005b) have indeed shown that a mass between 0.4 and 2.0
10$^{14}$ M$_\odot$ originating from field galaxy groups has fallen
onto the Coma cluster in a Hubble time. Assuming that all this matter
was constituted of galaxies, we then have a strict upper value of the
number of galaxies captured by Coma. This represents a maximum of
$\sim$4000 galaxies of 0.1 M$^*$ (using Mamon 2000).

Mamon (2000) gives an estimate in a Hubble time of the fusion number
of 0.1 M$^*$ galaxies with lower mass objects as a function of
distance to the cluster center. For simplicity we will use the values
given at 0.05, 0.1, 0.2 and 0.3 virial radii.
Multiplying these numbers by the maximal number of infalling galaxies,
we then have estimates of the numbers of objects fainter than 0.1
M$^*$ accreted by 0.1 M$^*$ galaxies as a function of distance to the
cluster center. By comparing these numbers with the density of lacking
galaxies in the south relative to the north, we find that the proposed
merging process is only able to explain $\sim$25$\%$ of the lacking
galaxies at 0.05 virial radius, $\sim$5$\%$ at 0.1 virial radius and
less than 1$\%$ at 0.2 and 0.3 virial radius. Therefore this process
can have a significant effect close to the cluster center, but is
overall inefficient above 0.1 virial radius.

- Destruction of faint galaxies by tidal effects cannot act alone
because we do not detect any significant large scale diffuse light
source in the south (see Adami et al. 2005a), far from the dominant
cluster galaxies.  It could possibly act locally but is not sufficient
to explain the entire faint galaxy deficit. We also estimated the
possible number of faint galaxies removed by this process using Mamon
(2000) and Adami et al. (2005b). Counting the number of strong tidal
encounters, we found that this effect is at maximum 1/10th of the
previous process (merging of faint galaxies with brighter objects). It
is therefore negligible.

- A last explanation would be the inhibition of faint galaxy formation
in the south compared to the north. We know for example that faint 
galaxies can be formed by tidal encounters (e.g. Bournaud et al. 2003) inside 
the cluster. An example of such formed galaxies could be the blue low surface 
brightness galaxies of Adami et al. (2006b). This 
process could be inhibited by modest tidal forces induced by infalling 
galaxies. We show here a marginal overabundance of blue low surface 
brightness galaxies in the north subfields compared to the south. Assuming
the same overabundance for all faint galaxies, this can, however, only explain 
$\sim$10$\%$ of the galaxies lacking in the south.

In conclusion, merging of faint galaxies with brighter objects and faint 
galaxy formation inhibition in the south can only explain part of the south 
galaxy deficit compared to the north. Another explanation, that still has to 
be verified via numerical simulations, is an accumulation in the north of 
faint galaxies coming from the south.

\subsection{What happens around the dominant galaxies NGC~4874 and
NGC~4889?}

The LF steeply rises around NGC~4874 in I and R, is flatter in V and
quite flat in B. Moreover, most of the objects around NGC~4874 are
red, except at very faint magnitudes where blue and red objects have
the same contribution. We must find the process(es) explaining such
a behaviour.

We have already shown that a non negligible part of the objects around
NGC~4874 have colors consistent with those of GCs, i.e.  are red.
NGC~4874 is also probably the oldest dominant galaxy in the Coma
cluster (e.g. Adami et al. 2005b or Neumann et al. 2003), therefore
located at the bottom of the cluster potential well and affected by very
  strong environmental effects (e.g. Odell et al. 2002). This region is
very dense and populated by the oldest Coma cluster galaxies which
have probably already used up their gas reservoir, being now
quite quiescent from the star-forming point of view. This should
explain why they appear red and are not easily detected in the blue
band.

Around NGC~4889 the LF steeply rises in all the observed bands and
there are no specially red objects in this region. This is confirmed by
the low percentage of objects detected with colors consistent with
GCs. Moreover, if NGC~4889 has a recent history in the Coma cluster,
its satellite galaxies have not yet had time to stop their star
formation due to cluster environmental effects.

\subsection{Large scale dips in the Coma cluster luminosity functions}

We see several dips in the LFs across the Coma cluster field of view,
making the LFs clearly not purely Schechter-like. Are these dips real
or due to bias in the statistical subtraction method?  These dips
could come from empty field oversubtractions due to the existence of
structures not present along the Coma cluster line of sight. However,
they appear when using both the F02 or F10 comparison fields, so they
are probably real. They concern galaxies within a relatively
narrow magnitude range. We have no explanation for this, but it
probably indicates that the dips are linked to physical processes
involving only certain galaxy populations. 

For simplicity, we only discuss the dips seen in
Figs.~\ref{fig:fdlnord} and~\ref{fig:fdlsud}. As previously discussed,
at zero order a dip shows that galaxies have been removed from the
counts. In a simple model, these removals can be due either to the
disappearance of galaxies (through merging or destruction visible
in all the bands) or to galaxy faintening (star forming activity
inhibition mainly visible in the blue bands). We note that
a brief star forming activity enhancement during the merging or the
destruction is also possible, inducing a bump in the blue bands
and a dip in the red bands.

We chose to first search for dips in the B or V bands. Later,
we checked if these dips were also visible in the R and I bands.

We mainly detected two dips visible both in the north and south
fields: D1 and D2. We also detected two dips visible in the north
field (DN3 and DN4) and three in the south field (DS3, DS4 and
DS5). The dip characteristics are summarized in
Table~\ref{tab:dips}. 

\begin{table}
\caption{Dips seen at large scale across the Coma field of view.}
\begin{tabular}{llll}
\hline
 & Magnitude & Origin \\
\hline
D1 & B$\sim$15.75 & Bright elliptical mergings \\
D2 & B$\sim$17.5 & Disappearance of red elliptical-like galaxies  \\
   &             & Potentially detected in Andreon $\&$ Pell\'o (2000) \\
DN3 & B$\sim$19.25 & Disappearance of spiral galaxies   \\
DN4 & B$\sim$20.5 & Late type spiral star-formation inhibition \\
DS3 & B$\sim$18.5 & Disappearance of bright spiral galaxies \\
    &             & Potentially detected in Lobo et al. (1997) \\
DS4 & B$\sim$21.5 & ? \\
DS5 & V$\sim$22 & Disappearance of faint spiral galaxies \\
\hline
\end{tabular}
\label{tab:dips}
\end{table}

The results suggest that the southern region is probably subject to
the global disappearance of galaxies and not to simple star-forming
inhibition (because all the dips observed in the south are
visible in all the bands), while the northern region shows both
processes.

\section{CMR and the Coma cluster history}

It is remarkable that most of the faint Coma cluster galaxies overall
follow the bright galaxy RS. This means that a large fraction of the
faint Coma cluster galaxies have evolutionary paths similar to bright
cluster galaxies in terms of metallicities. However, if we look more
carefully at the Coma cluster galaxy distribution in the CMR, we see
blue and red objects close to the bright galaxy RS.

\subsection{Blue objects in the CMR}

These objects, detected nearly over the whole Coma cluster field
(Fig.~\ref{fig:mosaic}) except perhaps around NGC~4874 and not 
always significantly in the south have colors
typical of late type galaxies suggesting that they still have a
significant star-formation activity. As we already discussed, they are 
only marginally explained by GCs, and are probably galaxies.

We suggest that part of these objects can be recently formed galaxies in 
the Coma cluster. This is supported by the fact that we have
detected a large population of faint (R$<$21) and low surface
brightness galaxies in the Coma cluster with atypically
blue colors (Adami et al. 2006b). We suggested in this paper that
these objects originated from external material expelled from brighter
late type galaxies. Significant concentrations of these faint blue 
low surface brightness galaxies are detected in several places across 
the Coma cluster: mainly in the western part of the cluster and around 
NGC~4889. They also have the right B-R colors and the right R magnitudes 
to explain at least part of the blue objects visible in 
Fig.~\ref{fig:mosaic}. However, concentrations of blue Coma cluster 
objects are also detected where there is no significant concentration 
of faint blue low surface brightness galaxies. These galaxies are 
therefore not the only explanation for our blue objects.

\subsection{Red objects in the CMR}

Contrary to the blue galaxies, faint red objects close to the RS can
be at least partially explained by GCs. However we also detect red
objects which are too bright to be GCs, mainly in the south area and
around NGC~4874. Their colors suggest that they have a low
star-formation activity. They could also be similar to the
nucleated dwarfs studied by Rakos $\&$ Schombert (2004) that were
found to be redder than bright elliptical galaxies.

\subsection{Slope of the CMR}

We computed a slope of $-0.045$ for the RS bright early type galaxies
in A06a. This slope is very similar to the CMR slope (for all
available galaxy types) in the south regions. This is a way to show
that the galaxy populations in the south fields are mainly similar to
bright red early type galaxies. In the north, the negative slope is
steeper. The universal metallicity-mass relation found by Terlevich et
al. (2001) for the bright galaxies is then not valid for the whole
faint galaxy population.  This expresses the predominance of blue
objects that followed different evolutionary paths in the north fields
compared to bright ellipticals. More precise estimates in terms
of ages and metallicities are difficult because we do not know which
part of the observed characteristics are coming from internal galaxy
physics and which from external processes due to environmental
effects. We will therefore not push this discussion further.

\section{Conclusions}

In this paper, we have analyzed the deep galaxy LF in the B, V, R and
I bands and the faint B-R/R CMR for the Coma cluster. We applied a
single Schechter law to model our LFs. In order to sample possible
environmental effects on the LF, we derived twenty such functions in
10$\times$10~arcmin$^2$ areas. We found a dichotomy between steeply
rising LFs in the north-northeast Coma region, which appear to be due
to red/early type galaxies at bright magnitudes, and to blue/late type
objects at very faint magnitudes, and much flatter LFs in the
south-southwest region, both for all objects and for blue/late type or
red/early type galaxies. This result is valid for all photometric bands and
all observation epochs.

We also found that the LFs differ around the main Coma cluster
galaxies and groups. LFs in the NGC~4889 field are steeply rising in
all photometric bands while around NGC~4874, they are rising
moderately in B and V and steeply in R and I, with a maximum reached
around R$\sim$23 and I$\sim$22.5. Around NGC~4911, the LF is moderately
rising in B and V up to B$\sim$23.75 and V$\sim$23.5 and oscillating
and rising up to R$\sim$23 and over the whole I magnitude range.

We have discussed the observed shapes of the LFs in various regions in
the framework of the building up process that we have proposed in a
previous paper (Adami et al.  2005b). In this scenario, several
infalls onto the Coma cluster are taking place, coming from three
directions: northeast-east, northwest-west and south-west. NGC~4874 is
the original cluster dominant galaxy, while NGC~4889 experienced a
later capture. How does this match our observations?

The GC expelling scenario explains quite well the high fraction of GCs
found close to NGC~4874.

The south-southwest regions showing quite flat LFs are also those
which are directly located on the main field infall directions (in
particular the infalling NGC~4839 group). The observed paucity of
faint galaxies in the southern region is only partially explained 
by the fact that faint galaxies may have been swallowed by relatively 
bright galaxies on their way to the cluster center. Rather, north counts 
could be enhanced by these south infalls to explain their steeply rising 
LFs (the north part has a LF slope significantly different from the field,
e.g. Ilbert et al. 2005).  However, we can note that faint galaxies are 
blue/late type while bright galaxies are red/early types in the north. This 
suggests that these blue and faint galaxies could also partially be 
created in the cluster from material expelled from the envelopes of 
spiral-like parent galaxies.

We have detected several dips in the LFs in different areas of the
cluster and shown that some of them are probably due to mergings/fusions, 
while others are probably due to the inhibition of star-formation activity 
by cluster environmental effects.

Finally, we have also clearly detected a RS down to R=23.5,
close to the bright galaxy RS computed in A06a. This shows that most
of the faint Coma cluster galaxies have probably followed the same
evolutionary path as the bright ellipticals in terms of
mass-metallicity relation. However, part of the faint cluster 
population followed a different evolution on the blue and red sides. We
suggest that blue objects present over the whole Coma cluster field
are partially due to recent galaxy formation from material expelled
from late type galaxies. It is particularly interesting to note that the
CMR extends over such a large range of magnitudes, thus making it a
very useful tool for several purposes, such as selecting galaxies with
a high probability of belonging to a cluster from imaging data only.

In a near future, we intend to compute and discuss the Coma cluster
luminosity functions based on photometric redshifts as soon as we will
have acquired the  U band data lacking for our survey (see A06a).

\begin{acknowledgements}
The authors thank the referee for useful and constructive comments.
The authors are grateful to the CFHT and Terapix teams for their help
and to the French PNG, CNRS for financial support. The authors thank
B. Lanzoni and G. Mamon for useful discussions and for providing their
numerical simulations. 
\end{acknowledgements}

\end{document}